\documentclass[11pt]{article}

\usepackage{amssymb}

\newcommand{\mbf}[1]{\mbox{\boldmath ${#1}$}}

\newcommand{\ee}[1]{{\rm e}^{#1}}
\newcommand{\ii}{{\rm i}}
\newcommand{\dd}{{\rm d}}

\def\Dirac{{D\!\!\!\!/\,}} 

\newcommand{\eq}{\begin{equation}}
\newcommand{\eqend}{\end{equation}}
\newcommand{\eqa}{\begin{eqnarray}}
\newcommand{\nonueqa}{\begin{eqnarray*}}
\newcommand{\eqaend}{\end{eqnarray}}
\newcommand{\nonueqaend}{\end{eqnarray*}}
\newcommand{\nonu}{\nonumber \\ \nopagebreak}
\newcommand{\bma}[1]{\begin{array}{#1}}
\newcommand{\ema}{\end{array}}
\newcommand{\bc}{\begin{center}}
\newcommand{\ec}{\end{center}}

\newcommand{\pfaff}{{\rm Pfaff}}
\newcommand{\erf}{{\rm erf}}

\newcommand{\sine}{{\sf K}}
\newcommand{\sinec}{{\sf I}}
\newcommand{\sinecc}{{\mbf I}}
\newcommand{\sineccs}{{\bf I}}

\renewcommand{\thefootnote}{\fnsymbol{footnote}}
\newcommand{\newsection}{\setcounter{equation}{0}\section}

\def\appendix#1{\addtocounter{section}{1}\setcounter{equation}{0}
\renewcommand{\thesection}{\Alph{section}}
\section*{Appendix \thesection\protect\indent \parbox[t]{11.715cm} {#1}}
\addcontentsline{toc}{section}{Appendix \thesection\ \ \ #1} }

\newcommand{\complex}{{\mathbb C}} 
\newcommand{\zed}{{\mathbb Z}} 
\newcommand{\nat}{{\mathbb N}} 
\newcommand{\real}{{\mathbb R}} 
\newcommand{\symp}{{\mathbb J}} 
\newcommand{\id}{{1\!\!1}} 
\def\Dirac{{D\!\!\!\!/\,}} 

\newif\ifold             \oldtrue            \def\new{\oldfalse}

\def\nn{\nonumber}

\newcommand{\Tr}[1]{\:{\rm Tr}\,#1}

\def\e{{\,\rm e}\,}

\def\be{\begin{equation}}
\def\ee{\end{equation}}
\def\bea{\begin{eqnarray}}
\def\eea{\end{eqnarray}}
\def\bd{\begin{displaymath}}
\def\ed{\end{displaymath}}

\newcommand{\beq}{\begin{eqnarray}}
\newcommand{\eeq}{\end{eqnarray}}

\makeatletter
\newdimen\normalarrayskip              
\newdimen\minarrayskip                 
\normalarrayskip\baselineskip
\minarrayskip\jot
\newif\ifold             \oldtrue            \def\new{\oldfalse}
\def\arraymode{\ifold\relax\else\displaystyle\fi} 
\def\@arrayskip{\ifold\baselineskip\z@\lineskip\z@
     \else
     \baselineskip\minarrayskip\lineskip2\minarrayskip\fi}
\def\@arrayclassz{\ifcase \@lastchclass \@acolampacol \or
\@ampacol \or \or \or \@addamp \or
   \@acolampacol \or \@firstampfalse \@acol \fi
\edef\@preamble{\@preamble
  \ifcase \@chnum
     \hfil$\relax\arraymode\@sharp$\hfil
     \or $\relax\arraymode\@sharp$\hfil
     \or \hfil$\relax\arraymode\@sharp$\fi}}
\def\@array[#1]#2{\setbox\@arstrutbox=\hbox{\vrule
     height\arraystretch \ht\strutbox
     depth\arraystretch \dp\strutbox
     width\z@}\@mkpream{#2}\edef\@preamble{\halign \noexpand\@halignto
\bgroup \tabskip\z@ \@arstrut \@preamble \tabskip\z@ \cr}%
\let\@startpbox\@@startpbox \let\@endpbox\@@endpbox
  \if #1t\vtop \else \if#1b\vbox \else \vcenter \fi\fi
  \bgroup \let\par\relax
  \let\@sharp##\let\protect\relax
  \@arrayskip\@preamble}
\makeatother

\begin{document}
\begin{titlepage}
\begin{flushright}

\baselineskip=12pt

HWM--05--05\\
EMPG--05--06\\
hep--th/0504202\\
\hfill{ }\\
April 2005
\end{flushright}

\begin{center}

\baselineskip=24pt

\vspace{2cm}

{\Large\bf Finite Volume Gauge Theory Partition Functions\\ in Three
Dimensions}

\baselineskip=14pt

\vspace{2cm}

{\bf Richard J. Szabo}
\\[4mm]
{\it Department of Mathematics\\ School of Mathematical and Computer
  Sciences\\ Heriot-Watt University\\ Colin Maclaurin Building,
  Riccarton, Edinburgh EH14 4AS, U.K.}
\\{\tt R.J.Szabo@ma.hw.ac.uk}
\\[50mm]

\end{center}

\begin{abstract}

\baselineskip=12pt

We determine the fermion mass dependence of Euclidean finite volume partition
functions for three-dimensional QCD in the $\epsilon$-regime directly
from the effective field theory of the pseudo-Goldstone modes by using
zero-dimensional non-linear $\sigma$-models. New results are given for an
arbitrary number of flavours in all three cases of complex,
pseudo-real and real fermions, extending some previous
considerations based on random matrix theory. They are used to
describe the microscopic spectral correlation functions and smallest
eigenvalue distributions of the QCD$_3$ Dirac operator, as well
as the corresponding massive spectral sum rules.

\end{abstract}

\end{titlepage}
\setcounter{page}{2}

\newpage

\renewcommand{\thefootnote}{\arabic{footnote}} \setcounter{footnote}{0}

\newsection{Introduction and Summary \label{Intro}}

The $\epsilon$-regime of quantum chromodynamics in four spacetime
dimensions (QCD$_4$)~\cite{GL1,LS1} provides one of
the few examples wherein exact nonperturbative results can be derived
from QCD. It corresponds to the infrared sector of the Euclidean Dirac
operator spectrum which is related to the mechanism of chiral symmetry
breaking. The $\epsilon$-expansion of chiral perturbation theory in
this case, which describes the dynamics of the Goldstone modes~\cite{BG1}, is
taken with respect to the small quantity
$\epsilon=L^{-1}$ where $L$ is the linear size of the given spacetime
volume. If $L$ is much larger than the QCD scale, then spontaneously
broken chiral symmetry dictates the interactions and the leading
contributions come from the zero-momentum pseudo-Goldstone
fields. This regime represents a domain wherein a detailed
understanding can be achieved of the lowest-lying Dirac operator
eigenvalues, which dominate the physical low-energy observables, and
in which numerical computations in lattice QCD are
possible~\cite{GHLW1}. Most of the recent understanding of the
$\epsilon$-regime of QCD comes from its relation to universal random
matrix theory results~\cite{SV1,ADMN1}.

In this paper we will study the infrared behaviour of the Dirac
operator spectrum in three-dimensional QCD (QCD$_3$). In this case the
analog of chiral symmetry breaking is played by the spontaneous
breaking of flavour symmetry~\cite{Pisarski1}. The finite volume gauge
theory is connected to three universality classes in random matrix
theory, the unitary ensemble (fundamental fermions with
$N_c\geq3$ colours)~\cite{VZ1,HV1}, the orthogonal ensemble
(fundamental fermions with colour group $SU(2)$)~\cite{Magnea1} and
the symplectic ensemble (adjoint fermions with colour group
$SU(N_c)$)~\cite{Magnea2}. The random matrix theory
representation of the effective QCD$_3$ partition function in the
$\epsilon$-regime has lead to a number of conjectures regarding the
derivation of the spectral properties of the three-dimensional Dirac
operator from joint eigenvalue probability distributions in the
pertinent matrix model~\cite{VZ1,NS1}. The patterns of flavour
symmetry breaking in all three instances have been
identified~\cite{VZ1,Magnea1,Magnea2} and
supported by Monte Carlo simulations~\cite{DHKM1}, and the microscopic
spectral correlation functions have been
calculated~\cite{ADMN1},\cite{DamNish}--\cite{NN1}.

To derive nonperturbative analytic results for the smallest eigenvalues of the
Dirac operator it is necessary to have a complete proof that the
universal random matrix theory results coincide with the low-energy
limit of the effective field theory. From the field
theoretic point of view~\cite{LS1} these observables should be computable
entirely within the framework of finite volume partition functions
which are generating functionals for the order parameter for flavour
symmetry breaking~\cite{AD2}. For QCD$_3$ this problem has only been thoroughly
investigated for the unitary ensemble by using the supersymmetric
formulation of partially quenched effective Lagrangians~\cite{me}, the replica
limit~\cite{ADDV1}, and the relationship~\cite{ADS1} between finite
volume partition functions and the $\tau$-function of an underlying
integrable KP hierarchy~\cite{AD1,BMM1}. The
goal of this paper is to derive finite volume
partition functions in all three cases mentioned above and to use them
to explore spectral characteristics of the QCD$_3$ Dirac operator
directly in the low-energy effective field theory.

In all three ensembles the effective field theory is described by a
non-linear $\sigma$-model in a maximally symmetric compact
space~\cite{VZ1,Magnea1,Magnea2}. These target spaces fall into the
classical Cartan classification of symmetric spaces, and confirm part
of the Zirnbauer classification of random matrix universality classes
in terms of Riemannian symmetric superspaces~\cite{Zirn1}. This
enables us to exploit the well-known geometrical properties of these
spaces~\cite{Helg1,Helg2} (see~\cite{CM1} for a recent review in the
context of random matrix theory) and hence explore the differences between the
three classes of fermions in QCD$_3$. For example, we will find that
the finite volume partition functions for adjoint fermions are
drastically different in
functional form from those for complex and pseudo-real fermions, and this
difference can be understood from the form of the Riemannian metrics
on the corresponding symmetric spaces. A similar geometric feature
also exhibits the difference between even and odd numbers of quark
flavour components. In the latter cases the theory is plagued by a
severe sign problem which is rather delicate from the effective field
theory point of view~\cite{VZ1}. We derive new results in all three
instances for the finite volume partition functions with odd numbers of
flavours. In the case of the unitary ensemble, we use them to derive a
new simplified expression for the microscopic spectral correlators,
extending the known effective field theory results for the
even-flavoured case~\cite{me}. We also study some of the spectral sum
rules which constrain the eigenvalues of the Dirac
operator~\cite{VZ1,LS1,Dam1}, and briefly describe its smallest
eigenvalue distribution functions~\cite{Damsmall}.

On a technical level, we present a complete and thorough analysis only
for the case of complex fermions. In the cases of pseodo-real and real
fermions, we complete the derivations only in the cases of completely
degenerate mass eigenvalues, as there are some algebraic obstructions
which prevent us from obtaining closed analytical expressions for the
finite volume partition functions for arbitrary configurations of quark
masses. Nevertheless, we present a very general algebraic description
of the differences and similarities between the three universality
classes, and also of the distinction between the even and odd
flavoured cases, based on a remnant
discrete subgroup of the flavour symmetry acting on the mass
parameters of the gauge theory. This residual symmetry is described in
general in Section~2, where we also review various aspects of the
non-linear $\sigma$-models and describe the geometry of their
symmetric target spaces in a general setting that applies also to
other symmetric space $\sigma$-models such as those encountered in
QCD$_4$. Sections~3--5 then use this general formalism
to analyse in detail the three classes of fermions in turn. In
addition to obtaining new results in the odd-flavoured cases, the main
new predictions of the effective field theory approach concern the
(equal mass) partition functions for adjoint fermions, which have a
far more intricate functional structure than the other ensembles and
which involve functions usually encountered in QCD$_4$~\cite{LS1,GW1},
not in QCD$_3$.

\newsection{Finite Volume Partition Functions \label{Finitevol}}

The Euclidean Dirac operator governing the minimal coupling of quarks
in the QCD$_3$ action is given by
\beq
\ii\,\Dirac=\ii\,\sigma^\mu\,\left(\partial_\mu+\ii\,A_\mu\right) \ ,
\label{DiracQCD3def}\eeq
where $\sigma^\mu$, $\mu=1,2,3$ are the Pauli spin matrices
\beq
\sigma^1=\pmatrix{0&1\cr1&0\cr} \ , ~~ \sigma^2=\pmatrix{0&-\ii\cr\ii&0\cr} \ ,
{}~~ \sigma^3=\pmatrix{1&0\cr0&-1\cr} \ ,
\label{Pauli}\eeq
and $A_\mu$ are gauge fields of the local colour group
$SU(N_c)$, $N_c\geq2$. When multi-colour fermion fields are coupled to
this operator, the three-dimensional field theory also possesses a
global flavour symmetry described by rotations of the quark field
components via elements of some Lie group $G$. If one introduces quark
masses in equal and opposite pairs, then the fermion determinant is
positive definite and the quantum field theory is anomaly
free~\cite{VZ1}. The addition of such mass terms preserves
three-dimensional spacetime parity~\cite{Christ1} but breaks the flavour
symmetry, and thereby permits one to analyse the possibility of spontaneous
flavour symmetry breaking in QCD$_3$. In this section we will
construct the general low-energy effective field theory of the
associated pseudo-Goldstone bosons in the ergodic $\epsilon$-regime wherein
exact, nonperturbative information about the spectrum of the QCD$_3$
Dirac operator (\ref{DiracQCD3def}) can be obtained.

\subsection{Non-Linear $\mbf\sigma$-Models of QCD$_{\bf3}$
  \label{NLsigmaQCD3}}

We are interested in the mechanism of flavour symmetry breaking in
three-dimensional QCD, which we consider defined at fixed ultraviolet
cutoff $\Lambda$, while the quantum field theory in the infrared is
regulated by a three-volume $V$. The effective field theory for the
low-momentum modes of the Nambu-Goldstone bosons can be derived by using the
three-dimensional version of chiral perturbation theory~\cite{BG1}. It is
consistent with the flavour symmetry of the original quantum field
theory represented through transformations of the quark fields by the
Lie group $G$. In the ergodic regime, the effective Lagrangian in
Euclidean space is given by
\beq
L_{\rm eff}=\mbox{$\frac{f_\pi^2}4$}
\,\Tr\Bigl(\partial_\mu{\cal U}\,\partial^\mu
{\cal U}^*\Bigr)-\mbox{$\frac{\Sigma}2$}\,\Tr\Bigl[{\cal M}\left
({\cal U}+{\cal U}^*\right)\Bigr]+O\left({\cal M}^2\right) \ ,
\label{Leff}\eeq
where $f_\pi$ is the decay constant of the pseudo-Goldstone modes,
$\Sigma$ is the infinite-volume quark-antiquark condensate which is the order
parameter for flavour symmetry breaking, and $\cal M$ is the
$N_f\times N_f$ mass matrix induced by integrating out the $N_f\geq1$
flavours of fermion fields in the QCD$_3$ partition function. When the
number of fermion flavours $N_f=2n$ is even, so that $n$ is the
number of four-component spinors, we assume that the eigenvalues of
$\cal M$ occur in parity-conjugate pairs $(m_i,-m_i)$, $m_i\geq0$,
$i=1,\dots,n$. When $N_f=2n+1$ is odd, there is a single unpaired quark of
mass $m_0\geq0$ which breaks parity symmetry explicitly when $m_0>0$
and radiatively when $m_0=0$~\cite{Redlich1}. The higher order terms
in (\ref{Leff}) represent contributions involving gluons and confined
quarks. The symbol $\Tr$, in the representations that we shall use in
this paper, will always correspond to an ordinary $N_f\times N_f$
matrix trace. The Nambu-Goldstone field $\cal U$ lives in the
appropriate vacuum manifold for the given symmetry breaking. We will
use the Dyson index $\beta=1,2,4$, which is borrowed from random matrix theory
terminology~\cite{Mehta1}. It labels the anti-unitary symmetries of the QCD$_3$
Dirac operator which in turn depends on the types of fermions that are
present in the original field theory~\cite{HV1}.

The Goldstone manifold is a coset space
\beq
{\cal G}_\beta(N_f)~=~G\,/\,H_\Gamma \ ,
\label{Goldstonebeta}\eeq
where $H_\Gamma$ is the stability subgroup of the original flavour
symmetry group $G$ which leaves fixed an $N_f\times N_f$ parity matrix
$\Gamma$ representing an involutive automorphism of $G$ that reflects
the parity symmetry of the original field theory. The fields in
(\ref{Leff}) may then be parametrized as
\beq
{\cal U}=U\,\Gamma\,U^{\rm t}
\label{calUpar}\eeq
with $U\in{\cal G}_\beta(N_f)$. In (\ref{calUpar}) the superscript t
denotes the appropriate conjugation involution on the Lie group $G$,
given by $U^{\rm t}=U^\dag$ for $\beta=2$ and $U^{\rm t}=U^\top$ for
$\beta=1,4$.

Given the non-linear $\sigma$-model (\ref{Leff}), we
would now like to study the region where the zero-momentum mode of $U$
dominates. This is the $\epsilon$-regime in which the linear dimension of the
system is much smaller than the Compton wavelength of the Nambu-Goldstone
bosons. The spacetime integration over the effective Lagrangian
(\ref{Leff}) then produces an overall volume factor~$V$, and the
partition function of the non-linear $\sigma$-model simplifies to a
finite-dimensional coset integral. The region of interest is obtained
by simultaneously taking the thermodynamic limit $V\to\infty$ and the
chiral limit ${\cal M}\to{\mbf0}_{N_f}$ of massless quarks, with the
rescaled masses $M=V\,\Sigma\,{\cal M}$ finite. The effective field
theory of the Goldstone bosons is then the correct description of the
dynamics below the scale set by the mass gap. Because of the tuning of
the quark masses to zero, the higher-order corrections to this
effective theory are exponentially small in the mass gap and come from
integrating out the heavy states.

The quantum field theory is thus adequately described by the finite
volume partition function~\cite{VZ1,Magnea1,Magnea2}
\bea
Z_\beta^{(N_f)}(M)&=&\int\limits_{{\cal G}_\beta(N_f)}\,\dd\nu_\beta(U)~
\Big[\exp~{\rm Re}\,\Tr\left(M\,U\,\Gamma\,U^{\rm t}\right)\nonu
&&\qquad\qquad\qquad\quad+\,\mbox{$\frac12$}\,
\left(1-(-1)^{q_{N_f}}\right)\,(-1)^{N_\Dirac}~
\exp~{\rm Re}\,\Tr\left(-M\,U\,\Gamma\,U^{\rm t}\right)\Big]\nonu &&
\label{ZMgen}\eea
where $q_{N_f}$ is the even/odd congruence class of the flavour number
$N_f$ given by $q_{2n}=0$, $q_{2n+1}=1$, and $N_\Dirac$ is the number of Dirac
operator eigenvalues (defined, for example, in a lattice
regularization or in random matrix theory).The second line of
(\ref{ZMgen}) only contributes when there is an odd number of fermion
flavours. It arises from the fact that, generally, the QCD$_3$
partition function changes by a phase $(-1)^{N_f\,N_\Dirac}$ under a
parity transformation of the quark masses. When $N_f=2n+1$ is odd, the
fermion determinant is not positive definite and its sign determines
two gauge inequivalent vacuum states which both have to be included in
the low-energy limit of the QCD$_3$ partition function in the
$\epsilon$-regime~\cite{VZ1}.  We will always ignore irrelevant numerical
constants (independent of $M$) in the evaluation
of (\ref{ZMgen}), as they will not contribute to any physical
quantities and can be simply absorbed into the normalization of the
partition function. The measure $\dd\nu_\beta(U)$ is inherited from
the invariant Haar measure for integration over the Lie group $G$. It
will be constructed in the next subsection.

The low-energy effective partition function of the zero-dimensional
$\sigma$-model (\ref{ZMgen}) contains sufficient information to
completely constrain the low-energy Dirac operator spectrum. For instance, by
matching exact results from random matrix theory with the low-energy
effective field theory (\ref{ZMgen}), it is possible to express the
microscopic limit of the spectral $k$-point correlation functions
\beq
\varrho^{(N_f)}_\beta(\lambda_1,\dots,\lambda_k;{\cal M})
=\left\langle\,
\prod_{l=1}^k\Tr\,\delta(\lambda_l-\ii\,\Dirac+\ii\,\mathcal{M})
\right\rangle_\beta
\label{kptDirac2}\eeq
of the QCD$_3$ Dirac operator (\ref{DiracQCD3def}) in terms of a ratio of
finite volume partition functions, namely (\ref{ZMgen}) and one which
involves $\beta k$ additional species of valence quarks of imaginary
masses~$\ii\,\lambda_l$~\cite{AD2,AD1}. In (\ref{kptDirac2}) the
average is taken over $SU(N_c)$ gauge field configurations weighted by
the three-dimensional Yang-Mills action. Both (\ref{ZMgen}) and
the microscopic limit $\rho_2^{(N_f)}$ of (\ref{kptDirac2}) depend only on the
eigenvalues $\pm\,\mu_i$, $i=1,\dots,n$ (and $\mu_0$ when $N_f$ is odd)
of the scaled mass matrix $M$.

For $\beta=2,4$ one explicitly has

\vbox{\bea
\rho^{(N_f)}_\beta(\zeta_1,\dots,\zeta_k;M)&=&
C_\beta^{(k)}\,\prod_{l=1}^k\left(\,\ii\,\zeta_l+\mu_0\right)^{q_{N_f}}
\,\prod_{i=1}^n\left(\zeta_l^2+\mu_i^2
\right)~\prod_{m<p}\left|\zeta_m^2-\zeta_p^2\right|^\beta\nonu&&
\times\,\frac{Z_\beta^{(N_f+\beta k)}(M,\overbrace{
\ii\,\zeta_1,\dots,\ii\,\zeta_1}^\beta,\dots,\overbrace{
\ii\,\zeta_k,\dots,\ii\,\zeta_k}^\beta\,)}{Z_\beta^{(N_f)}(M)}
\label{kptexplbeta}\eea}\noindent
where $\zeta_l=V\,\Sigma\,\lambda_l$ are the unfolded Dirac operator
eigenvalues, and the proportionality constant $C_\beta^{(k)}$ is fixed
by the matching condition between the confluent limit of the
microscopic correlators and the macroscopic correlation functions at
the spectral origin as
\beq
\lim_{\zeta_1,\dots,\zeta_k\to\infty}\,\rho_\beta^{(N_f)}
(\zeta_1,\dots,\zeta_k;M)=\frac{\varrho_\beta^{(N_f)}
(\,\overbrace{0,\dots,0}^k\,;
{\cal M})}{(V\,\Sigma)^k}=\left(\frac1\pi\right)^k \ .
\label{match}\eeq
For $\beta=1,2,4$, the finite volume partition functions may be used to
compute the normalized hole probability $E_\beta^{(N_f)}(\zeta;M)$,
i.e. the probability that the interval $[-\zeta,\zeta]\subset\real$ is
free of Dirac operator eigenvalues. One has the
formula~\cite{Damsmall}
\beq
E_\beta^{(N_f)}(\zeta;M)=\e^{-\frac14\,\zeta^2}~\frac{Z_\beta^{(N_f)}
\left(\,\sqrt{M^2+\zeta^2}\,\right)}{Z_\beta^{(N_f)}(M)} \ ,
\label{holeprob}\eeq
from which the smallest eigenvalue distribution is given by
\beq
P_\beta^{(N_f)}(\zeta;M)=-\frac{\partial E_\beta^{(N_f)}(\zeta;M)}
{\partial\zeta} \ .
\label{smallprob}\eeq

Furthermore, by matching (\ref{ZMgen}) with the
formal expansion of the full QCD$_3$ partition function
\beq
{\cal Z}_\beta^{(N_f)}(\mathcal{M})=\left\langle\,\prod_{i=1}^n
\det\left(\Dirac^2+m_i^2\right)^{p_\beta}~
\det\left(\Dirac+\ii\,m_0\right)^{q_{N_f}\,p_\beta}\right\rangle_\beta
\label{QCD3partfnfull}\eeq
with $p_1=p_2=1$ and $p_4=\frac12$, one can express the mass-dependent
susceptibilities in terms of derivatives of the finite
volume partition functions with respect to one or more quark
mass eigenvalues~\cite{Dam1}. This leads to a set of massive spectral
sum rules once (\ref{ZMgen}) is known explicitly. The simplest such sum rule
expresses the mass-dependent condensate for the flavour symmetry
breaking in the case of equal quark masses as
\beq
\left\langle\,\sum_{\zeta>0}\frac1{\zeta^2+\mu^2}\right\rangle_\beta=
\frac1{2N_f\,\mu}\,\frac{\partial\ln Z_\beta^{(N_f)}(\mu,\dots,\mu)}
{\partial\mu} \ .
\label{sumrulegen}\eeq
In the massless case $\mu\to0$, the left-hand side of
(\ref{sumrulegen}) can be computed directly from the coset integral
(\ref{ZMgen})~\cite{VZ1}--\cite{Magnea2} and is given as a function of
the Dyson index $\beta$ by
\beq
\left\langle\,\sum_{\zeta>0}\frac1{\zeta^2}\right\rangle_\beta=
\frac{N_f-q_{N_f}}{2(N_f-1)\,\left(\frac{2(N_f-q_{N_f})}\beta+1\right)}
\label{masslesssumrule}\eeq
for even values of $N_\Dirac$. The formula (\ref{masslesssumrule})
generalizes the one given in~\cite{Magnea1,Magnea2,HN1,Magnea3} to
include the case of odd $N_f$. For $\beta=2$ it was first obtained
in~\cite{VZ1}. Note that for an odd number of Dirac
operator eigenvalues, the finite volume partition function
(\ref{ZMgen}) is not positive and vanishes when the unpaired fermion
is massless, i.e. $\mu_0=0$~\cite{Christ1}. It is therefore not
suitable to describe physical quantities such as chiral spectral sum
rules, while it can appear radiatively~\cite{AD2}.

\subsection{Parametrization of the Goldstone Manifold
  \label{GoldstonePar}}

We will now describe how to evaluate the non-linear $\sigma$-model
partition functions (\ref{ZMgen}). For this, we shall first review some
elementary Lie theory that will be of use to us and which will fix
some notation. We assume that the quark flavour symmetry is
represented by a compact, connected, simple Lie group $G$, which
we take to be realized by $N_f\times N_f$ invertible matrices $U$. Let
$\mathfrak t$ be a Cartan subalgebra of the Lie algebra $\mathfrak g$
of $G$, so that $T=\exp(\,\ii\,{\mathfrak t}\,)$ is a maximal torus of
$G$. Any element $X\in\mathfrak g$ can be taken to lie in the Cartan
subalgebra ${\mathfrak t}$ without loss of generality, since this can
always be achieved via rotations by elements of $G$ if necessary.

After performing such a rotation in the given Cartan decomposition of
$G$, the Jacobian for the change of Haar integration measure is given
by a Weyl determinant $\Delta_G(X)$ of $G$ which is determined through
the product
\beq
\Delta_G(X)=\prod_{\alpha>0}\,(\alpha,X) \ ,
\label{pXdef}\eeq
where $(\alpha,X)$ are the positive roots of $G$ evaluated on the Cartan
element $X$. Its most useful form is obtained by choosing an orthonormal basis
$\vec e_i$, $i=1,\dots,\dim T$ of vectors in weight space, $\vec
e_i\cdot\vec e_j=\delta_{ij}$. Then the root vectors $\alpha$ can be
identified with dual elements
\beq
\alpha^\vee=\sum_{i=1}^{\dim T}\alpha_i\,\vec e_i \ ,
\label{alphastar}\eeq
while
\beq
X=\sum_{i=1}^{\dim T}x_i\,H_i
\label{XHiexpand}\eeq
where $H_i$ are the generators of $\mathfrak t$ in the matrix basis
 corresponding to the orthonormal weights. By identifying
 (\ref{XHiexpand}) with the weight vector $x^\vee=\sum_ix_i\,\vec e_i$,
 the pairing in (\ref{pXdef}) may then be written as the inner product
\beq
(\alpha,X)=\alpha^\vee\cdot x^\vee=\sum_{i=1}^{\dim T}\alpha_i\,x_i \ .
\label{alphaX}\eeq

With $U=\exp(\,\ii\,X)$, the suitably normalized invariant Haar measure
on the Lie group $G$ is given by
\beq
[\dd U]=\prod_{i=1}^{\dim T}\dd x_i~\prod_{\alpha>0}\sinh^2(\alpha,\ln U) \ .
\label{HaarG}\eeq
Its infinitesimal form induces a measure on the Lie algebra
$\mathfrak g$ as
\beq
[\dd X]=\prod_{i=1}^{\dim T}\dd x_i~\Delta_G(X)^2 \ .
\label{Haarg}\eeq
For the Lie groups that we shall deal with in this paper, the Cartan
subalgebra $\mathfrak t$ is represented in the basis (\ref{XHiexpand}) by
$N_f\times N_f$ diagonal matrices.

The residual flavour symmetry after such a transformation is given by the
  discrete Weyl group $W_G$ of $G$ which consists of the inequivalent
  transformations on ${\mathfrak t}\to{\mathfrak t}$ given by the
  adjoint actions
\beq
X~\longmapsto~X^w=w\,X\,w^{-1} \ , ~~ w\in G \ .
\label{hatwdef}\eeq
Geometrically, $W_G$ is the group of automorphisms of the root lattice of
$G$ corresponding to Weyl reflections. To each $w\in W_G$ we may
associate a sign factor ${\rm sgn}(w)=\pm\,1$ which is the even/odd
$\zed_2$-parity of the Weyl element according to its action
(\ref{hatwdef}) on the diagonal elements of the Cartan matrices. For the
partition functions that will be of interest to us in this paper, the
relevant group theoretical data are summarized in
Table~\ref{Gtable}.

\begin{table}
\begin{center}
\begin{tabular}{|c||c|c|c|c|}
\hline $G$ & Positive Weights & $H_i$ & Dynkin Diagram & Weyl Group\\
\hline\hline $U(r)$ & $\vec e_i-\vec e_j \ ,$ & $E_{i,i} \ ,$ &
\begin{picture}(70.00,5.00)
\thinlines
\put(3.00,2.00){\circle{6.00}}
\put(6.00,2.00){\line(1,0){8.00}}
\put(17.00,2.00){\circle{6.00}}
\put(20.00,2.00){\line(1,0){8.00}}
\put(31.00,2.00){\circle{6.00}}
\put(36.00,2.00){$\dots$}
\put(57.00,2.00){\circle{6.00}}
\put(60.00,2.00){\line(1,0){8.00}}
\put(71.00,2.00){\circle{6.00}}
\end{picture}
& $S_{r}$\\$\scriptstyle\beta=2$ & $1\leq i<j\leq r$ & $1\leq i\leq r$ & & \\
\hline $Sp(2r)$ & $\vec e_i\pm \vec e_j \ , ~~ 1\leq i<j\leq r$ &
$E_{i,i}-E_{i+r,i+r} \ ,$ &
\begin{picture}(70.00,5.00)
\thinlines
\put(3.00,2.00){\circle{6.00}}
\put(3.00,5.00){\line(1,0){14.00}}
\put(3.00,-1.00){\line(1,0){14.00}}
\put(17.00,2.00){\circle*{6.00}}
\put(20.00,2.00){\line(1,0){8.00}}
\put(31.00,2.00){\circle*{6.00}}
\put(36.00,2.00){$\dots$}
\put(57.00,2.00){\circle*{6.00}}
\put(60.00,2.00){\line(1,0){8.00}}
\put(71.00,2.00){\circle*{6.00}}
\end{picture}
& $S_r\ltimes(\zed_2)^r$\\$\scriptstyle\beta=1$ & $2\vec e_i \ , ~~
 1\leq i\leq r$ & $1\leq i\leq r$
 &  & \\
\hline $SO(2r+1)$ & $\vec e_i\pm \vec e_j \ , ~~ 1\leq i<j\leq r$ &
$E_{i,i}-E_{i+r,i+r} \ ,$ &
\begin{picture}(70.00,5.00)
\thinlines
\put(3.00,2.00){\circle*{6.00}}
\put(3.00,5.00){\line(1,0){14.00}}
\put(3.00,-1.00){\line(1,0){14.00}}
\put(17.00,2.00){\circle{6.00}}
\put(20.00,2.00){\line(1,0){8.00}}
\put(31.00,2.00){\circle{6.00}}
\put(36.00,2.00){$\dots$}
\put(57.00,2.00){\circle{6.00}}
\put(60.00,2.00){\line(1,0){8.00}}
\put(71.00,2.00){\circle{6.00}}
\end{picture}
 & $S_r\ltimes(\zed_2)^r$\\$\scriptstyle\beta=4$ & $\vec e_i \ , ~~
 1\leq i\leq r$ & $1\leq i\leq r$
& & \\ \hline
$SO(2r)$ & $\vec e_i\pm \vec e_j \ ,$ & $E_{i,i}-E_{i+r,i+r} \ ,$
&
\begin{picture}(70.00,25.00)
\thinlines
\put(3.00,2.00){\circle{6.00}}
\put(6.00,2.00){\line(1,0){8.00}}
\put(17.00,2.00){\circle{6.00}}
\put(22.00,2.00){$\dots$}
\put(43.00,2.00){\circle{6.00}}
\put(46.00,2.00){\line(1,0){8.00}}
\put(57.00,2.00){\circle{6.00}}
\put(57.00,5.00){\line(0,1){8.00}}
\put(57.00,16.00){\circle{6.00}}
\put(60.00,2.00){\line(1,0){8.00}}
\put(71.00,2.00){\circle{6.00}}
\end{picture}
& $S_r\ltimes(\zed_2)^{r-1}$\\$\scriptstyle\beta=4$ & $1\leq i<j\leq r$
& $1\leq i\leq r$ & & \\
\hline
\end{tabular}
\end{center}
\caption{\baselineskip=12pt {\it The flavour symmetry groups $G$ of rank
    $r=\dim T\geq2$ and their corresponding Dyson index~$\beta$. The
vectors $\vec e_i$, $i=1,\dots,r$ form an orthonormal basis of weight space,
while $E_{\mu,\nu}$ is a corresponding orthonormal basis of matrix units
$(E_{\mu,\nu})_{\lambda\rho}=\delta_{\mu\lambda}\,\delta_{\nu\rho}$. The
Dynkin diagrams all have $r$ nodes. The symmetric group $S_r$ permutes the
eigenvalues $x_i$ of a given Cartan element $X\in{\mathfrak t}$, while the
cyclic group $\zed_2$ reflects them.}}
\label{Gtable}\end{table}

We are now ready to construct the measure $\dd\nu_\beta(U)$ for
integration over the Goldstone manifold (\ref{Goldstonebeta}). For
this, we shall introduce polar coordinates on the coset
space ${\cal G}_\beta(N_f)$~\cite{Helg1}, which in each case will be
an irreducible, maximally symmetric compact space falling into the
Cartan classification. Let
\beq
\mathfrak{g}=\mathfrak{h}^{~}_\Gamma\oplus{\mathfrak h}^\perp_\Gamma
\label{Cartandecomp}\eeq
be a Cartan decomposition of the simple algebra $\mathfrak g$ with
respect to the stability subgroup
$H_\Gamma=\exp(\,\ii\,\mathfrak{h}_\Gamma^{~})$, with
$\mathfrak{h}_\Gamma^\perp=\mathfrak{g}\ominus\mathfrak{h}^{~}_\Gamma$ the
orthogonal complement of $\mathfrak{h}^{~}_\Gamma$ in $\mathfrak g$ with
respect to the Cartan-Killing form. The condition for
(\ref{Goldstonebeta}) to be a symmetric space is then
\beq
\bigl[\mathfrak{h}^{~}_\Gamma\,,\,\mathfrak{h}^{~}_\Gamma
\bigr]\subset\mathfrak{h}^{~}_\Gamma \
, ~~ \bigl[\mathfrak{h}^{~}_\Gamma\,,\,\mathfrak{h}^\perp_\Gamma
\bigr]\subset\mathfrak{h}^\perp_\Gamma \ , ~~
\bigl[\mathfrak{h}_\Gamma^\perp\,,\,\mathfrak{h}_\Gamma^\perp
\bigr]\subset\mathfrak{h}^{~}_\Gamma \ .
\label{symmspaceconds}\eeq

Let ${\mathfrak t}_\Gamma$ be a maximal abelian subalgebra in the subspace
$\mathfrak{h}_\Gamma^\perp$, and let $\mathcal{N}_{{\mathfrak
    t}_\Gamma}$ be the normalizer subgroup of ${\mathfrak t}_\Gamma$
in $H_\Gamma$ with $\mathcal{C}_{{\mathfrak t}_\Gamma}\subset
\mathcal{N}_{{\mathfrak t}_\Gamma}$ its centralizer subgroup. As
$H_\Gamma$ is a compact symmetric subgroup of the flavour symmetry
group $G$, every coset element $U\in{\cal G}_\beta(N_f)$ can be written by
means of the adjoint representation of the stationary subgroup
$H_\Gamma$ as~\cite{Helg1}
\beq
U=V\,R\,V^{-1} \ ,
\label{radialcoordgen}\eeq
where $V\in H_\Gamma/\mathcal{C}_{{\mathfrak t}_\Gamma}$ while $R\in
  T_\Gamma=\exp(\,\ii\,{\mathfrak t}_\Gamma)$ up to transformations by
  the adjoint actions of elements of the factor group
  $\mathcal{N}_{{\mathfrak t}_\Gamma}/\mathcal{C}_{{\mathfrak
      t}_\Gamma}$, which can be identified
  with the Weyl group of the restricted root system on the symmetric
  space ${\cal G}_\beta(N_f)$. This new root lattice differs from that
  inherited from the original group $G$ if the Cartan subalgebra
  $\mathfrak t$ is a subset of $\mathfrak{h}^{~}_\Gamma$, and it is
  defined by completing the maximal abelian subalgebra ${\mathfrak
  t}_\Gamma^{~}\subset{\mathfrak h}_\Gamma^\perp$ with the generators in
  $\mathfrak{h}^{~}_\Gamma$ that commute with ${\mathfrak t}_\Gamma^{~}$ to
give a
  new representation of $\mathfrak t$ that lies partly in
  $\mathfrak{h}_\Gamma^\perp$. The matrix $R$ in
  (\ref{radialcoordgen}) is the radial coordinate of the point
  $U\in{\cal G}_\beta(N_f)$, while $V$ is its angular coordinate. This
  defines a foliation of the Goldstone manifold (\ref{Goldstonebeta})
  by conjugacy classes under the adjoint action of the stability group
  $H_\Gamma$ of the symmetric space.

The decomposition (\ref{radialcoordgen}) means
  that every matrix $U\in{\cal G}_\beta(N_f)$ can be diagonalized by a
  similarity transformation in the subgroup $H_\Gamma$, and the radial
  coordinates $\{r_i\}$ are exactly the set of
  eigenvalues of $U$. Using the Haar measure (\ref{HaarG}) on the Lie
  group $G$, the corresponding Jacobian for the change of integration
  measure by (\ref{radialcoordgen}) can be computed by standard
  techniques and one finds~\cite{Helg2,ZH1}
\beq
\dd\nu_\beta(U)=[\dd V]~\prod_{i=1}^{\dim T_\Gamma}\dd
r_i~\prod_{\alpha_\Gamma^{~}>0}\,
\Bigl|\sinh(\alpha_\Gamma^{~},\ln R)\Bigr|^{m_{\alpha_\Gamma^{~}}} \ ,
\label{Jacradial}\eeq
where $\ln R=\ii\,\sum_ir_i\,H_i$ with
  $r_i\in[0,\frac\pi2]$ for $i=1,\dots,\dim T_\Gamma$. The second product
  in (\ref{Jacradial}) goes over positive roots of the {\it restricted} root
  lattice on ${\cal G}_\beta(N_f)$, and $m_{\alpha_\Gamma^{~}}$ is the
  multiplicity of the root $\alpha_\Gamma^{~}$ in the given Cartan
  decomposition, i.e. the dimension of the subspace of raising
  operators corresponding to the root $\alpha_\Gamma^{~}$ in the algebra
  $\mathfrak{h}^{~}_\Gamma$~\cite{Helg2}. One can further exploit the
  ambiguity in the choice of radial coordinates $R$ labelled by the
  restricted Weyl group $\mathcal{N}_{{\mathfrak
t}_\Gamma}/\mathcal{C}_{{\mathfrak
      t}_\Gamma}$ to alternatively parametrize the radial
  decomposition (\ref{radialcoordgen}) in a way more tailored to the
  evaluation of (\ref{ZMgen}). The particular details of this
  parametrization will depend in general on the random matrix ensemble
  in which we are working. As we will see, the radial flavour symmetry
  group $\mathcal{N}_{{\mathfrak t}_\Gamma}/\mathcal{C}_{{\mathfrak
      t}_\Gamma}$ of the Goldstone manifold naturally distinguishes
  the cases $\beta=1,2$ from $\beta=4$, and in all cases provides a
  nice algebraic distinction between the cases of an even or odd
  number of flavours $N_f$.

\subsection{The Harish-Chandra Formula \label{HarishChandra}}

The computation of (\ref{ZMgen}) can be compared with the
Harish-Chandra formula which computes the {\it group}
integral~\cite{HC}
\beq
\int\limits_G\,[\dd U]~\exp\,\Tr\left(X\,U\,Y\,U^{-1}\right)
=\frac1{\Delta_G(X)\,\Delta_G(Y)}
\,\sum_{w\in W_G}{\rm sgn}(w)~\exp\,\Tr\left(X\,Y^w\right)
\label{HCformula}\eeq
for $X,Y\in{\mathfrak t}$. It expresses the group integral as a finite sum
over terms, one for each element of the Weyl group, and it is intimately
related to the Fourier transform of the Weyl character formula for the
flavour symmetry group $G$. Its main characteristic is
that the integration in (\ref{HCformula}) may be rewritten as an integral over
a coadjoint orbit, which is a symmetric space to which the Duistermaat-Heckman
localization formula may be applied~\cite{mebook}. The Weyl group can be
described as the factor group $W_G=\mathcal{N}_G/T$, where
$\mathcal{N}_G$ is the normalizer
subgroup of fixed points of the left action of $T$ on the orbit $G/T$. In other
words, the right-hand side of (\ref{HCformula}) is the stationary phase
approximation to the group integral, and the Harish-Chandra formula states that
in this case the approximation is {\it exact}. For this to be true it is
important that the matrices $X$ and $Y$ live in the Lie algebra
$\mathfrak{g}$ of $G$.

\newsection{Fundamental Fermions\label{FF}}

The simplest instance is when the quarks are described by complex
fundamental Dirac fermion fields with $N_c\geq3$ colours. Although
these models have been the most extensively studied ones thus far, it
will be instructive for later comparisons and simply for completeness
to rederive the finite volume partition functions in this case using
the formalism of the previous section. In matrix model language it
corresponds to the unitary ensemble $\beta=2$. The global flavour
symmetry group is then $U(N_f)$, while the mass and parity matrices
are given in a suitable basis of the vector space $\complex^{N_f}$ by
\bea
M&=&\left\{\matrix{\sigma^3\otimes\pmatrix{\scriptstyle\mu_1& & \cr
      &\scriptstyle\ddots& \cr & &\scriptstyle\mu_n} &,& N_f=2n \ ,
\cr \pmatrix{\mu_0& \cr &\sigma^3\otimes\pmatrix{\scriptstyle\mu_1& & \cr
      &\scriptstyle\ddots& \cr & &\scriptstyle\mu_n}} &,& N_f=2n+1 \ ,
} \right. \nonumber\\[15pt]
\Gamma&=&\left\{\matrix{\sigma^3\otimes\id_n &,& N_f=2n \ , \cr
\pmatrix{1& \cr &\sigma^3\otimes\id_n} &,&N_f=2n+1 \ . } \right.
\label{MGamma2}\eea
The mass matrix explicitly breaks the flavour symmetry, and its
pairing into quark masses of opposite sign preserves three-dimensional
spacetime parity. The Goldstone manifolds are~\cite{VZ1}
\bea
{\cal G}_2(2n)&=&U(2n)\,/\,U(n)\times U(n) \ , \label{calG2}\\
{\cal G}_2(2n+1)&=&U(2n+1)\,/\,U(n+1)\times U(n)
\label{G22n1}\eea
of real dimensions $\dim{\cal G}_2(2n)=2n^2$ and $\dim{\cal
  G}_2(2n+1)=2n\,(n+1)$.

\subsection{Partition Functions\label{PFFund}}

\subsubsection*{\mbf{N_f=1}}

The case of a single massive quark corresponds to formally setting $n=0$ in
(\ref{G22n1}). In this case there is no flavour symmetry breaking, the
Goldstone manifold is a point, and there are no massless propagating
degrees of freedom. The finite volume partition function is then
trivially obtained from (\ref{ZMgen}) in the form
\beq
\begin{tabular}{|c|}\hline\\
$
Z_2^{(1)}(\mu)=\left\{\matrix{\cosh\mu &,& N_\Dirac~{\rm even} \cr
\sinh\mu &,& N_\Dirac~{\rm odd}} \right.
$\\\\\hline\end{tabular}
\label{Z21mu}\eeq

\subsubsection*{\mbf{N_f=2n}}

The restricted root lattice of the symmetric space (\ref{calG2}) is
  the root lattice $C_n$ of the symplectic group $Sp(2n)$ (see
  Table~\ref{Gtable}), with multiplicities $m_{\vec e_i\pm\vec
  e_j}=2$, $m_{2\vec e_i}=1$ for $i,j=1,\dots,n$,
$i<j$~\cite{Helg1,CM1}. The coset space measure (\ref{Jacradial}) for
the case $\beta=2$ and an even number of fermion flavours is thereby
given as
\beq
\dd\nu_2(U)=[\dd V]~\prod_{i=1}^n\dd r_i~\sin2r_i~
\prod_{i<j}\sin^2(r_i-r_j)\,\sin^2(r_i+r_j)
\label{Jacradial2}\eeq
with $r_i\in[0,\frac\pi2]$. By using the trigonometric identity
\beq
2\sin(r_i-r_j)\,\sin(r_i+r_j)=\cos2r_i-\cos2r_j
\label{trigid}\eeq
and defining $\lambda_i=\cos2r_i\in[-1,1]$, we may write
(\ref{Jacradial2}) in the simpler form
\beq
\dd\nu_2(U)=\frac{[\dd V]}{2^{n^2}}~\prod_{i=1}^n\dd\lambda_i~\Delta(
\lambda_1,\dots,\lambda_n)^2 \ .
\label{dnu2Ulambdai}\eeq
For any $X\in\mathfrak{u}(n)$ represented as a diagonal $n\times n$
matrix with entries $x_i\in\real$, we have introduced its Vandermonde
determinant which is the Weyl determinant (\ref{pXdef},\ref{alphaX})
for the unitary group $G=U(n)$ given by
\beq
\Delta_{U(n)}(X)=\prod_{i<j}\,(x_i-x_j)=\Delta(x_1,\dots,x_n) \ .
\label{Vandermonde}\eeq

In the polar decomposition (\ref{radialcoordgen}), we define the
$n\times n$ diagonal matrix
\beq
\rho=\pmatrix{\scriptstyle{r_1}& & \cr
 &\scriptstyle{\ddots}& \cr & &\scriptstyle{r_n}}
\label{rhodef}\eeq
and write the radial coordinates as
\beq
R=\exp\left(\,\ii\,\sigma^1\otimes\rho\right)
\label{Rbeta2}\eeq
which, since $\{\sigma^i,\sigma^j\}=2\delta^{ij}$,
has the requisite property that it anticommutes with the parity matrix
in (\ref{MGamma2}) as
\beq
R\,\Gamma=\Gamma\,R^{-1} \ .
\label{RGammaprop}\eeq
We decompose the angular degrees of freedom $V\in U(n)\times U(n)$ as
\beq
V=\pmatrix{V_+& \cr &V_-}
\label{Vbeta2}\eeq
with $V_\pm\in U(n)$, so that they commute with the parity matrix as
\beq
V\,\Gamma=\Gamma\,V \ .
\label{VGammaprop2}\eeq
On substituting (\ref{radialcoordgen}) into the finite volume
partition function (\ref{ZMgen}), and by using the identities
(\ref{Rbeta2})--(\ref{VGammaprop2}) along with
$\cos(\sigma^1\otimes\eta)=\id_2\otimes\cos\eta$ for any
$\eta\in{\mathfrak u}(n)$, we may thereby write the action of the
non-linear $\sigma$-model in this case as

\vbox{\bea
{\rm Re}\,\Tr\left(M\,U\,\Gamma\,U^\dag\,\right)&=&
\mbox{$\frac12$}\,\Tr\left[M\,\Gamma\,V\,\left(R^2+R^{-2}\right)\,
V^{-1}\right]\nn\\&&{~~~~}^{~~}_{~~}\nn\\&=&
\Tr\left[M\,\Gamma\,V\,\cos\left(2\sigma^1\otimes\rho\right)\,
V^{-1}\right]\nn\\&&{~~~~}^{~~}_{~~}\nn\\&=&\Tr\left[
\pmatrix{\scriptstyle\mu_1& & \cr
 &\scriptstyle\ddots& \cr & &\scriptstyle\mu_n}\,V^{~}_+\,
\pmatrix{\scriptstyle\lambda_1& & \cr
 &\scriptstyle\ddots& \cr & &\scriptstyle\lambda_n}\,
V_+^{-1}\right]\nonu&&+\,\Tr\left[\pmatrix{\scriptstyle\mu_1& & \cr
 &\scriptstyle\ddots& \cr & &\scriptstyle\mu_n}\,V^{~}_-\,
\pmatrix{\scriptstyle\lambda_1& & \cr
 &\scriptstyle\ddots& \cr & &\scriptstyle\lambda_n}\,
V_-^{-1}\right] \ .
\label{sigmaaction2}\eea}

The resulting integrals over the two sets of angular coordinates
$V_\pm\in U(n)$ are identical and can be evaluated by applying the
Harish-Chandra formula (\ref{HCformula}) to the unitary group
$G=U(n)$ (In this case (\ref{HCformula}) is often refered to as the
Itzykson-Zuber formula~\cite{IZ} in the physics literature). The Weyl
group in this case is the symmetric group
$W_{U(n)}=S_n$ and it acts on $X$ by permuting its
eigenvalues, $(X^w)_i=(w\,X\,w^{-1})_i=x_{\hat w(i)}$, where $\hat
w\in S_n$ is a permutation and ${\rm sgn}(w)$ is its parity. In this
way the finite volume partition function (\ref{ZMgen}) acquires the
form
\beq
Z_2^{(2n)}(\mu_1,\dots,\mu_n)=\frac1{\Delta(\mu_1,\dots,\mu_n)^2}\,
\sum_{\hat w_\pm\in S_n}{\rm sgn}\left(\hat w_+\,\hat w_-\right)\,
\prod_{i=1}^n~\int\limits_{-1}^1\dd\lambda_i~\e^{\lambda_i(\mu_{
\hat w_+(i)}+\mu_{\hat w_-(i)})}
\label{Zmu2perm}\eeq
involving the unfolded masses $\mu_i$, $i=1,\dots,n$. Since the
integration measure and domain in (\ref{Zmu2perm}) are invariant under
permutations of the $\lambda_i$'s, we may reduce the double sum over
the Weyl group to a
{\it single} sum over the relative permutation $\hat w=\hat
w_+^{~}\,\hat w_-^{-1}$. The $\lambda_i$ integrals are
elementary and expressed in terms of the function
\beq
\sine(x)=\frac{\sinh x}x=\frac12\,\int\limits_{-1}^1\dd\lambda~
\e^{x\,\lambda}
\label{sinexdef}\eeq
which for purely imaginary argument defines the spectral sine-kernel
of the unitary ensemble~\cite{Mehta1}. The sum over $\hat w\in S_n$ defines a
determinant, and in this way one may finally arrive at the simple
$n\times n$ determinant formula
\beq
\begin{tabular}{|c|}\hline\\
$\displaystyle
Z_2^{(2n)}(\mu_1,\dots,\mu_n)=
\frac{\det\Bigl[\sine(\mu_i+\mu_j)\Bigr]_{i,j=1,\dots,n}}
{\Delta(\mu_1,\dots,\mu_n)^2}
$\\\\\hline\end{tabular}
\label{Z2M}\eeq
This is precisely the form of the finite volume partition function
that was obtained in~\cite{me} from considerations based on
partially-quenched effective field theories.

\subsubsection*{\mbf{N_f=2n+1}}

The restricted root lattice of the symmetric space (\ref{G22n1}) is
the non-reduced irreducible root lattice $BC_n$ given by the union of
the root systems $B_n$ and $C_n$ of the groups $SO(2n+1)$ and $Sp(2n)$
(see Table~\ref{Gtable}), with multiplicities $m_{\vec e_i\pm\vec
  e_j}=2$, $m_{2\vec e_i}=1$ as above, and $m_{\vec
  e_i}=2$~\cite{Helg1,CM1}. In this case the measure (\ref{Jacradial})
thus reads
\bea
\dd\nu_2(U)&=&[\dd V]~\prod_{i=1}^n\dd r_i~\sin2r_i\,\sin^2r_i~
\prod_{i<j}\sin^2(r_i-r_j)\,\sin^2(r_i+r_j)\nn\\ &=&
\frac{[\dd V]}{2^{n\,(n+1)}}~
\prod_{i=1}^n\dd\lambda_i~(1-\lambda_i)~\Delta(\lambda_1,\dots,
\lambda_n)^2 \ ,
\label{nu2measodd}\eea
where we have used the double angle identity
$\sin^2r_i=\frac12\,(1-\cos2r_i)$. With the definition (\ref{rhodef}),
the radial coordinate matrix
\beq
R=\pmatrix{1& \cr &\exp\left(\,\ii\,\sigma^1\otimes\rho\right)}
\label{radial2odd}\eeq
satisfies the requisite condition (\ref{RGammaprop}). The angular
matrices $V\in U(n+1)\times U(n)$ are decomposed as in (\ref{Vbeta2})
with $V_+\in U(n+1)$ and $V_-\in U(n)$, and the non-linear
$\sigma$-model action is now modified to
\beq
{\rm Re}\,\Tr\left(M\,U\,\Gamma\,U^\dag\,\right)&=&
\Tr\left[\pmatrix{\scriptstyle\mu_0& & & \cr &\scriptstyle\mu_1& & \cr
 & &\scriptstyle\ddots& \cr & & &\scriptstyle\mu_n}\,V^{~}_+\,
\pmatrix{\scriptstyle 1& & & \cr &\scriptstyle\lambda_1& & \cr
 & &\scriptstyle\ddots& \cr & & &\scriptstyle\lambda_n}\,
V_+^{-1}\right]\nonu&&+\,\Tr\left[\pmatrix{\scriptstyle\mu_1& & \cr
 &\scriptstyle\ddots& \cr & &\scriptstyle\mu_n}\,V^{~}_-\,
\pmatrix{\scriptstyle\lambda_1& & \cr
 &\scriptstyle\ddots& \cr & &\scriptstyle\lambda_n}\,
V_-^{-1}\right] \ .
\label{action2odd}\eea

The Harish-Chandra formula applied to the integrations over $V_+\in
U(n+1)$ and $V_-\in U(n)$ then yields
\bea
Z_2^{(2n+1)}(\mu_0,\mu_1,\dots,\mu_n)&=&\frac{\prod\limits_{i=1}^n
(\mu_i-\mu_0)^{-1}}{\Delta(\mu_1,\dots,\mu_n)^2}~\sum_{\hat w_+\in
S_{n+1}}~\sum_{\hat w_-\in S_n}{\rm sgn}\left(\hat w_+\right)~
{\rm sgn}\left(\hat w_-\right)~\e^{\mu_{\hat w_+(0)}} \nonu &&
\times\,\prod_{i=1}^n~\int\limits_{-1}^1\dd\lambda_i~
\e^{\lambda_i(\mu_{\hat w_+(i)}+\mu_{\hat w_-(i)})}~~+~~
(-1)^{N_\Dirac}~\left\{\stackrel{\scriptstyle\mu_0\to-\mu_0}
{\scriptstyle\mu_i\to-\mu_i}\right\} \ . \nonu &&
\label{Z22n1HC}\eea
Note how the extra factors of $(1-\lambda_i)$ in (\ref{nu2measodd}) have
nicely cancelled against the $(n+1)\times(n+1)$ Vandermonde
determinant $\Delta(1,\lambda_1,\dots,\lambda_n)$ coming from the
$V_+$ integral. In the sum over $\hat w_-\in S_n$, we may regard $S_n$
as the subgroup of the permutation group $S_{n+1}$ acting only on the
indices $i=1,\dots,n$ in $\{0,1,\dots,n\}$. Then as before we can
truncate the double Weyl group sum in (\ref{Z22n1HC}) to a single sum
over $\hat w=\hat w_+^{~}\,\hat w_-^{-1}\in S_{n+1}$ by extending
$\hat w_-^{~}$ to all of $S_{n+1}$ through the definition $\hat
w^{~}_-(0)=0$. After performing the $\lambda_i$ integrals, the finite volume
partition function is thus given by the simple $(n+1)\times(n+1)$
determinant formula
\beq
\begin{tabular}{|c|}\hline\\
$\displaystyle
Z_2^{(2n+1)}(\mu_0,\mu_1,\dots,\mu_n)=\frac{\prod\limits_{i=1}^n
(\mu_i-\mu_0)^{-1}}{\Delta(\mu_1,\dots,\mu_n)^2}\,\left(
\det\left[\matrix{\e^{\mu_0}&\e^{\mu_j}\cr\sine(\mu_0+\mu_i)&
\sine(\mu_i+\mu_j)}\right]_{i,j=1,\dots,n}\right.$\\[10pt]$
\qquad\qquad\qquad\qquad
\qquad+\left.(-1)^{n+N_\Dirac}~\det\left[\matrix{\e^{-\mu_0}&\e^{-\mu_j}\cr
\sine(\mu_0+\mu_i)&\sine(\mu_i+\mu_j)}\right]_{i,j=1,\dots,n}\right)
$\\\\\hline\end{tabular}
\label{Z2odd}\eeq
This is a new simplified expression for the finite volume partition
function of QCD$_3$ with an odd number of fundamental fermions. It can
be readily checked to coincide with previous expressions (given by
$(2n+1)\times(2n+1)$ determinant formulas~\cite{AD2,ADDV1}) for low
numbers of flavours.

\subsection{Correlation Functions}

For $\beta=2$, the $k$-level correlation functions can be expressed as
a ratio of partition functions with $N_f$ and $N_f+2k$ flavours as in
(\ref{kptexplbeta}). By using (\ref{Z2M}) one finds the scaled
correlators for an even number of quark flavours given by
\beq
\begin{tabular}{|c|}\hline\\
$\displaystyle
\rho_2^{(2n)}(\zeta_1,\dots,\zeta_k;\mu_1,\dots,\mu_n)=
\left(\frac1\pi\right)^k{}~\frac{
\det\left[\new{\begin{array}{cc}\sine(\mu_i+\mu_j)&
\sine(\mu_i+\ii\,\zeta_{l'})\\
\sine(\mu_j-\ii\,\zeta_l)&\sine(\ii\,\zeta_l-\ii\,\zeta_{l'})
\end{array}}\right]_{\stackrel{\scriptstyle i,j=1,\dots,n}{\scriptstyle
l,l'=1,\dots,k}}}{\new{\begin{array}{c}\det
\Bigl[\sine(\mu_i+\mu_j)\Bigr]_{i,j=1,\dots,n}\end{array}}}
$\\\\\hline\end{tabular}
\label{rhoskpointfinal}\eeq
where the matching condition (\ref{match}) in this case fixes
$C_2^{(k)}=\bigl(\frac1{2\pi}\bigr)^k$. This again agrees with the
field theoretic results of~\cite{me}. In the form (\ref{Z2M}), the expression
(\ref{kptexplbeta}), originally derived in the context of random
matrix theory~\cite{AD2}, is thereby naturally tied to the field theory
derivation of the spectral functions $\rho_2^{(2n)}$.

For an odd number of flavours one finds
\bea
&\begin{tabular}{|l|}\hline\\
$\scriptstyle{\rho_2^{(2n+1)}(\zeta_1,\dots,\zeta_k;\mu_0,
\mu_1,\dots,\mu_n)}$\\$\qquad\qquad\qquad=~
\frac{\det\left[\matrix{\scriptstyle\e^{\mu_0}&\scriptstyle
\e^{\mu_j}&\scriptstyle\e^{\ii\,\zeta_{l'}}\cr
\scriptstyle\sine(\mu_0+\mu_i)&\scriptstyle\sine(\mu_i+\mu_j)&
\scriptstyle\sine(\mu_i+\ii\,\zeta_{l'})\cr\scriptstyle
\sine(\mu_0-\ii\,\zeta_l)&
\scriptstyle\sine(\mu_j-\ii\,\zeta_l)&\scriptstyle
\sine(\ii\,\zeta_l-\ii\,\zeta_{l'})}
\right]_{\stackrel{\scriptstyle i,j=1,\dots,n}{\scriptstyle
l,l'=1,\dots,k}}~~+~~(-1)^{k+n+N_{\Dirac}}~\left\{\matrix{{\mu_0\to-\mu_0}
\atop{\mu_i\to-\mu_i} \cr \scriptstyle\zeta_l\to-\zeta_l}\right\}}
{(2\pi)^k~\det\left[\matrix{\scriptstyle
\e^{\mu_0}&\scriptstyle\e^{\mu_j}\cr\scriptstyle\sine(\mu_0+\mu_i)&
\scriptstyle\sine(\mu_i+\mu_j)}\right]_{i,j=1,\dots,n}~~+~~(-1)^{n+N_\Dirac}~
\left\{\stackrel{\scriptstyle\mu_0\to-\mu_0}{\scriptstyle
\mu_i\to-\mu_i}\right\}}
$\\\\\hline\end{tabular} \nonu &
\label{rhoskptodd}\eea
This is a new general expression for the spectral $k$-point
functions in the case of an odd number of quarks at $\beta=2$. It
coincides with the known expressions from random matrix
theory~\cite{Christ1} and the replica method approach to
QCD$_3$~\cite{ADS1}. In particular, the formula (\ref{rhoskptodd})
shows explicitly that the correlators are not even functions of the
scaled Dirac operator eigenvalues $\zeta_l$ for $M\neq\mbf
0_{2n+1}$~\cite{ADS1}. This expression also covers the case of a
single flavour $N_f=1$ ($n=0$), which can be obtained by omitting the
corresponding $n\times n$ blocks in the determinants above.

\subsection{Sum Rules\label{SRFund}}

We may work out the massive spectral sum rules (\ref{sumrulegen}) by
taking the equal mass limits in (\ref{Z2M}) and (\ref{Z2odd}). This
yields indeterminate forms which can be computed by regulating the
mass matrices in any way that removes the $n$-fold and $(n+1)$-fold
eigenvalue degeneracies, and then taking the degenerate limits
using l'H\^opital's rule. However, this is rather cumbersome to do in
practise and it is much simpler to go back and work directly with the
radial coset representation of the finite volume partition functions.

\subsubsection*{\mbf{N_f=2n}}

{}From (\ref{sigmaaction2}) we see that in the case of equal quark
masses $\mu_i=\mu~~\forall i=1,\dots,n$, the $\sigma$-model action is
independent of the angular degrees of freedom $V_\pm\in U(n)$ and the
low-energy dynamics is mediated entirely by the $n$ radial Goldstone
boson degrees of freedom. The
partition function (\ref{ZMgen}) is then simply given by
\beq
Z_2^{(2n)}(\mu,\dots,\mu)=\prod_{i=1}^n~\int\limits_{-1}^1\dd\lambda_i~
\e^{2\mu\,\lambda_i}~\Delta(\lambda_1,\dots,\lambda_n)^2 \ .
\label{Z2equalmu}\eeq
We now use the identity~\cite{Mehta1}
\beq
\Delta(\lambda_1,\dots,\lambda_n)=\det\left[\lambda_i^{j-1}
\right]_{i,j=1,\dots,n}
\label{Vandermondedet}\eeq
and expand the resulting two determinants in (\ref{Z2equalmu}) into sums
over permutations $\hat w_\pm\in S_n$. As in Section~\ref{PFFund},
the partition function depends only on the relative permutation $\hat
w=\hat w^{~}_+\,\hat w_-^{-1}\in S_n$, and the $\lambda_i$ integrals
are now expressed in terms of derivatives of the spectral kernel
defined by (\ref{sinexdef}) as
\beq
\sine^{(m)}(x)=\frac{\dd^m\sine(x)}{\dd
  x^m}=\frac12\,\int\limits_{-1}^1\dd\lambda~\lambda^m~\e^{x\,\lambda}
\ .
\label{Knxint}\eeq
In this way we arrive at the $n\times n$ determinant formula
\beq
Z_2^{(2n)}(\mu,\dots,\mu)=\det\Bigl[\sine^{(i+j-2)}(2\mu)
\Bigr]_{i,j=1,\dots,n} \ .
\label{Z2derivK}\eeq

We may present the partition function (\ref{Z2derivK}) in a somewhat
more explicit form by expressing the generic derivatives
(\ref{Knxint}) in terms of the sine-kernel $\sine(x)$ itself. Starting
from the integral representation (\ref{Knxint}) we write
\bea
\sine^{(m)}(x)&=&\frac12\,\int\limits_{-1}^1\dd\lambda~\e^{x\,\lambda}
\,\left(1-m\,\int\limits_\lambda^1\dd s~s^{m-1}\right)\nn\\&=&
\sine(x)-\frac m{2x}\,\int\limits_{-1}^1\dd s~s^{m-1}\,
\left(\e^{x\,s}-\e^{-x}\right) \ .
\label{Kmxints}\eea
This yields a simple linear inhomogeneous recursion relation for the
sequence of functions $\sine^{(m)}(x)$, $m\geq0$ in (\ref{Knxint})
given by
\beq
\sine^{(m)}(x)+\frac mx~\sine^{(m-1)}(x)=\sine(x)+\frac12\,
\Bigl(1-(-1)^m\Bigr)\left(\frac{\cosh x}x-\sine(x)\right) \ .
\label{Knrecursion}\eeq
It is straightforward to iterate the recurrence (\ref{Knrecursion}) in
$m$ and express the solution in terms of $\sine(x)$ and other elementary
functions of $x$, and in this way the equal mass partition function
(\ref{Z2derivK}) can be written as
\beq
\begin{tabular}{|c|}\hline\\
$\displaystyle
Z_2^{(2n)}(\mu,\dots,\mu)=\det\Bigl[\mbf K_{i+j-2}(2\mu)\Bigr]_{i,j=1,\dots,n}
$\\\\\hline\end{tabular}
\label{Z2detAmn}\eeq
where
\bea
\mbf K_m(x)&=&\frac12\,\sum_{k=0}^m
\frac{m!}{(m-k)!}\,\left(-\frac1x\right)^{k+1}\nonu&&\times\,\left[
\Bigl(1-(-1)^{m+k}\Bigr)\,\Bigl(\sinh x-\cosh x\Bigr)-2\sinh x\right]
\ .
\label{Knxij}\eea

{}From (\ref{Z2detAmn}) one may now proceed to
derive the massive spectral sum rules (\ref{sumrulegen}). For the
first few even numbers of quark flavours we find
\bea
&\begin{tabular}{|l|}\hline\\
$\left.\Biggl\langle\,\sum\limits_{\zeta>0}\frac1{\zeta^2+\mu^2}\Biggr\rangle_2
{}~\right|_{N_f=2}=\frac{2\mu\,\coth2\mu-1}{4\mu^2}
$\\${~~~~}^{~~}_{~~}$\\
$\left.\Biggl\langle\,\sum\limits_{\zeta>0}\frac1{\zeta^2+\mu^2}\Biggr\rangle_2
{}~\right|_{N_f=4}=\frac1{4\mu^2}\,\frac{2\sinh^22\mu-\mu\,\sinh4\mu
-4\mu^2}{4\mu^2-\sinh^22\mu}$\\${~~~~}^{~~}_{~~}$\\
$\left.\Biggl\langle\,\sum\limits_{\zeta>0}\frac1{\zeta^2+\mu^2}\Biggr\rangle_2
{}~\right|_{N_f=6}=\frac1{12\mu^2}\,\frac{2\mu\left(16\mu^4+60\mu^2+3
\right)\cosh2\mu
-28\mu^2\left(4\mu^2+3\right)\sinh2\mu-6\mu\,\cosh^32\mu+9\sinh^32\mu}
{4\mu^2\left(4\mu^2+3\right)\sinh2\mu-16\mu^3\,\cosh2\mu-\sinh^32\mu}
$\\\\\hline\end{tabular}\nonu&
\label{Nf2beta2}\eea
These all agree with (\ref{masslesssumrule}) in the limit
$\mu\to0$. The first two sum rules in (\ref{Nf2beta2}) were also
obtained in~\cite{DamNish}.

\subsubsection*{\mbf{N_f=2n+1}}

{}From (\ref{nu2measodd}) and (\ref{action2odd}) it follows that in the
equal mass limit $\mu_0=\mu_i=\mu~~\forall i=1,\dots,n$ the finite volume
partition function (\ref{ZMgen}) is given by
\beq
Z_2^{(2n+1)}(\mu,\dots,\mu)=\e^\mu~\prod_{i=1}^n~\int\limits_{-1}^1
\dd\lambda_i~(1-\lambda_i)~\e^{2\mu\,\lambda_i}~\Delta(\lambda_1,
\dots,\lambda_n)^2 ~~+~~ (-1)^{N_\Dirac}~\bigl\{\mu\to-\mu\bigr\} \ . \nonu
\label{Z2oddequal}\eeq
Proceeding exactly as above, and using the fact that the functions
(\ref{Knxij}) obey the reflection property $\mbf K_m(-x)=(-1)^m\,\mbf
K_m(x)$, we can write (\ref{Z2oddequal}) explicitly as the $n\times n$
determinant formula
\bea
&\begin{tabular}{|l|}\hline\\
$\displaystyle
Z_2^{(2n+1)}(\mu,\dots,\mu)=\e^\mu~\det\Bigl[\mbf K_{i+j-2}(2\mu)-
\mbf K_{i+j-1}(2\mu)\Bigr]_{i,j=1,\dots,n}$\\$\qquad\qquad\qquad\qquad\quad
+\,(-1)^{N_\Dirac}~\e^{-\mu}~\det\Bigl[
\mbf K_{i+j-2}(2\mu)+\mbf K_{i+j-1}(2\mu)\Bigr]_{i,j=1,\dots,n}
$\\\\\hline\end{tabular}
\label{Z2oddequalfinal}\eea
with the determinants omitted for the $n=0$ case as before. For an
even number $N_\Dirac$ of Dirac operator eigenvalues, from
(\ref{Z21mu}) and (\ref{Z2oddequalfinal}) we obtain the massive
spectral sum rules
\beq
\begin{tabular}{|l|}\hline\\
$\displaystyle
\left.\Biggl\langle\,\sum\limits_{\zeta>0}\frac1{\zeta^2+\mu^2}\Biggr\rangle_2
{}~\right|_{N_f=1}=\frac{\tanh\mu}{2\mu}
$\\${~~~~}^{~~}_{~~}$\\
$\displaystyle
\left.\Biggl\langle\,\sum\limits_{\zeta>0}\frac1{\zeta^2+\mu^2}\Biggr\rangle_2
{}~\right|_{N_f=3}=\frac1{6\mu^2}\,\frac{2\mu^2\,\coth\mu-\mu(1+3\cosh2\mu)
-2\sinh2\mu}{2\mu+\sinh2\mu}
$\\\\\hline\end{tabular}
\label{Nf2beta2odd}\eeq
which again agree with (\ref{masslesssumrule}) in the limit
$\mu\to0$.

\newsection{Pseudo-Real Fermions \label{Pseudo}}

The next simplest case is the parity-invariant Dirac operator
$\ii\,\Dirac$ acting on $N_c=2$ colours of Dirac quarks in the fundamental
representation of the $SU(N_c)$ gauge group. In contrast to the
previous case, the Dirac operator now possesses a special anti-unitary
symmetry, $[\mathfrak{C},\ii\,\Dirac]=0$, generated by an operator
$\mathfrak{C}$ with $\mathfrak{C}^2=\id$~\cite{HV1,Magnea1}. The
operator $\mathfrak{C}$ can be expressed as a combination of the usual
charge conjugation $C=\ii\,\sigma^2$ with
$C\,(\sigma^\mu)^*\,C^{-1}=\sigma^\mu$ for $\mu=1,2,3$, an
infinitesimal rotation by the matrix $\ii\,\sigma^2$ in colour space,
and complex conjugation. In a basis of spinors $|\psi_k\rangle$
wherein $\mathfrak{C}|\psi_k\rangle=|\psi_k\rangle$, the matrix elements
$\langle\psi_k|\,\ii\,\Dirac\,|\psi_l\rangle$ are all
real~\cite{Magnea1}. This basis is
obtained by augmenting the flavour components of the fermion fields by
the two-colour group, thereby enhancing the original $U(N_f)$ flavour
symmetry to a higher pseudo-real global flavour symmetry group
$Sp(2N_f)$. In matrix model language this corresponds to the
orthogonal ensemble $\beta=1$.

The group $USp(2n)$ acts on the vector space $\complex^{2n}$
preserving its canonical inner product as well as a symplectic
structure $\symp_n:\real^{2n}\to\real^{2n}$, i.e. a real antisymmetric
$2n\times2n$ matrix with
$\symp_n^2=-\id_{2n}^{~}$. In a suitable basis of $\complex^{2n}$, it
consists of unitary matrices $V$ satisfying
\beq
V\,\symp_n\,V^\top=\symp_n \ , ~~ \symp_n=\ii\,\sigma^2\otimes\id_n \ .
\label{symplgpdef}\eeq
With respect to the same basis, the mass and parity matrices are given
by
\bea
M&=&\left\{\matrix{\sigma^3\otimes\ii\,\sigma^2\otimes
\pmatrix{\scriptstyle\mu_1& & \cr &\scriptstyle
\ddots& \cr & &\scriptstyle\mu_n\cr} &,& N_f=2n \  , \cr
\pmatrix{\mu_0~\ii\,\sigma^2& \cr & \sigma^3\otimes\ii\,\sigma^2\otimes
\pmatrix{\scriptstyle\mu_1& & \cr &\scriptstyle
\ddots& \cr & &\scriptstyle\mu_n\cr}\cr} &,& N_f=2n+1 \ , \cr}
\right.\nonumber\\[15pt] \Gamma&=&\left\{\matrix{
\sigma^3\otimes\symp_n &,& N_f=2n \ , \cr
\pmatrix{\,\ii\,\sigma^2& \cr &\sigma^3\otimes\symp_n\cr} &,&
N_f=2n+1 \ . \cr} \right.
\label{massparsympl}\eea
The antisymmetric forms of these matrices follow from the Grassmann
nature of the quark spinor fields $\psi_k$ described above. Again the
mass matrices break the $USp(4n)$ and $USp(4n+2)$ flavour symmetries,
and the corresponding Goldstone manifolds are
\bea
\mathcal{G}_1(2n)&=&~USp(4n)\,/\,USp(2n)\times USp(2n) \ ,
\label{Goldstone1}\\
\mathcal{G}_1(2n+1)&=&~USp(4n+2)\,/\,USp(2n+2)\times USp(2n)
\label{Goldstone1odd}\eea
of real dimensions $\dim\mathcal{G}_1(2n)=4n^2$ and
$\dim\mathcal{G}_1(2n+1)=4n\,(n+1)$. As expected, the number of
Goldstone bosons here is exactly twice that of the previous section.

\subsection{Partition Functions\label{PFSymp}}

\subsubsection*{\mbf{N_f=2n}}

The restricted root lattice of the symmetric space (\ref{Goldstone1})
is again the root space $C_n$ of $Sp(2n)$, but now with multiplicities
$m_{\vec e_i\pm\vec e_j}=4$, $m_{2\vec e_i}=3$~\cite{Helg1,CM1}. The
corresponding
radial integration measure (\ref{Jacradial}) on $\mathcal{G}_1(2n)$ is
thus
\bea
\dd\nu_1(U)&=&[\dd V]~\prod_{i=1}^n\dd r_i~\sin^32r_i~
\prod_{i<j}\sin^4(r_i-r_j)\,\sin^4(r_i+r_j)\nonumber\\&=&
\frac{[\dd V]}{2^{n\,(2n-1)}}~
\prod_{i=1}^n\dd\lambda_i~\left(1-\lambda_i^2\right)~
\Delta(\lambda_1,\dots,\lambda_n)^4 \ ,
\label{dnu1U}\eea
where we have again defined $\lambda_i=\cos2r_i\in[-1,1]$. With $\rho$
defined as in (\ref{rhodef}), we parametrize the radial element $R$ as
\beq
R=\exp\left(\,\ii\,\sigma^1\otimes\xi\right)
\label{radialpar1}\eeq
where the matrix
\beq
\xi=\sigma^2\otimes\rho
\label{xidef}\eeq
commutes with the symplectic structure as
\beq
\xi\,\symp_n=\symp_n\,\xi \ .
\label{xisympcomm}\eeq
It follows that $R^\top=R^{-1}=R^\dag$, and the radial matrix
(\ref{radialpar1}) anticommutes with the parity matrix $\Gamma$ as in
(\ref{RGammaprop}). For the angular degrees of freedom $V\in
USp(2n)\times USp(2n)$ we again take the decomposition (\ref{Vbeta2})
with $V_\pm\in USp(2n)$, so that from (\ref{symplgpdef}) we have
\beq
V\,\Gamma\,V^\top=\Gamma \ .
\label{VGammabeta1}\eeq

Using these relations, along with the identity
$\cos(\sigma^2\otimes\eta)=\id_2\otimes\cos\eta$ for any
$\eta\in\mathfrak{u}(n)$, we can now write the action of the
non-linear $\sigma$-model (\ref{ZMgen}) similarly to
(\ref{sigmaaction2}) as
\bea
&&{\rm Re}\,\Tr\left(M\,U\,\Gamma\,U^\top\,\right)~=~
\mbox{$\frac12$}\,\Tr\left[\Gamma\,M\,V\,\left(R^2+R^{-2}
\right)\,V^{-1}\right]\nn\\&&{~~~~}^{~~}_{~~}\nn\\
&&~~~~~~~~~~~~~=~\Tr\left[\Gamma\,M\,V\,
(\id_4\otimes\cos2\rho)\,V^{-1}\right]\nn\\&&{~~~~}^{~~}_{~~}\nn\\
&&~~~~~~~~~~~~~=~
\Tr\left[\Biggl(\id_2\otimes\pmatrix{\scriptstyle\mu_1& & \cr
 &\scriptstyle\ddots& \cr & &\scriptstyle\mu_n}\Biggr)\,V^{~}_+\,
\Biggl(\id_2\otimes\pmatrix{\scriptstyle\lambda_1& & \cr
 &\scriptstyle\ddots& \cr & &\scriptstyle\lambda_n}\Biggr)\,
V_+^{-1}\right]\nonu&&~~~~~~~~~~~~~~~~~~+\,\Tr\left[\Biggl(\id_2\otimes
\pmatrix{\scriptstyle\mu_1& & \cr
 &\scriptstyle\ddots& \cr & &\scriptstyle\mu_n}\Biggr)\,V^{~}_-\,
\Biggl(\id_2\otimes\pmatrix{\scriptstyle\lambda_1& & \cr
 &\scriptstyle\ddots& \cr & &\scriptstyle\lambda_n}\Biggr)\,
V_-^{-1}\right] \ . \nonu&&
\label{sigmaaction1}\eea
The finite volume partition function (\ref{ZMgen}) thereby becomes
\beq
Z_1^{(2n)}(\mu_1,\dots,\mu_n)=\prod_{i=1}^n~\int\limits_{-1}^1
\dd\lambda_i~\left(1-\lambda_i^2\right)~\Delta(\lambda_1,\dots,\lambda_n)^4~
\Omega_1^{(n)}(\lambda_1,\dots,\lambda_n;\mu_1,\dots,\mu_n)^2
\label{Z1sympleven}\eeq
where
\bea
&&\Omega_1^{(n)}(\lambda_1,\dots,\lambda_n;\mu_1,\dots,\mu_n)\nonu && \qquad=
\int\limits_{Sp(2n)}[\dd V]~\exp\,\Tr\left[\Biggl(\id_2\otimes
\pmatrix{\scriptstyle\mu_1& & \cr
 &\scriptstyle\ddots& \cr & &\scriptstyle\mu_n}\Biggr)\,V\,
\Biggl(\id_2\otimes\pmatrix{\scriptstyle\lambda_1& & \cr
 &\scriptstyle\ddots& \cr & &\scriptstyle\lambda_n}\Biggr)\,
V^{-1}\right] \ . \nonu &&
\label{Sp2nintunknown}\eea
It is not at all clear at this stage how to proceed with the
calculation. The Lie algebra $\mathfrak{sp}(2n)$ consists of
$2n\times2n$ matrices $X$ obeying the equation
$X\,\symp_n+\symp_n\,X^\top=0$. The diagonal source matrices in
(\ref{Sp2nintunknown}) are {\it not} elements of this Lie algebra, nor
are the original mass and parity matrices in (\ref{massparsympl})
elements of $\mathfrak{sp}(4n)$. Thus the Harish-Chandra formula
(\ref{HCformula}) cannot be immediately applied to the angular
integral (\ref{Sp2nintunknown}). For the same reason this integral
cannot be evaluated by using character expansion techniques for the
symplectic group $Sp(2n)$~\cite{Bal1}. It would be interesting to determine the
corresponding analytical continuation of the Harish-Chandra formula to
this case, in analogy with that done in the unitary case $G=U(n)$~\cite{BW1}
which is based on the fact that the representation theories of the Lie
groups $U(n)$ and $GL(n,\complex)$ are essentially the same. Note that
the integral $\Omega_1^{(1)}(\lambda;\mu)$ can be computed by means of
the Harish-Chandra formula for $SU(2)\cong Sp(2)$.

In~\cite{NN1} the finite volume partition function in this case
is evaluated by applying a formalism based on skew-orthogonal
polynomials in the orthogonal ensemble of random matrix theory,
resulting in the Pfaffian expression
\beq
Z_1^{(2n)}(\mu_1,\dots,\mu_n)=\frac{{\rm Pfaff}\Bigl[\sine'
\left(\tilde\mu_a-\tilde\mu_b\right)\Bigr]_{a,b=1,\dots,2n}}
{\prod\limits_{i=1}^n\mu_i~\Delta\left(\mu_1^2,\dots,\mu_n^2
\right)^2}
\label{Z12nRMT}\eeq
where $\tilde\mu_{2i-1}=-\mu_i$, $\tilde\mu_{2i}=\mu_i$ for
$i=1,\dots,n$. Using the permutation expansion of the Pfaffian, the
integral representation (\ref{Knxint}) and by comparing
(\ref{Z12nRMT}) with (\ref{Z1sympleven}), it is possible to write a
conjectural formula for the integral (\ref{Sp2nintunknown}). There
are several qualitative features of the expression (\ref{Z12nRMT}) that make
its derivation directly in field theory appear feasible. First of all,
the denominator of (\ref{Z12nRMT}) can be expressed in terms of the
square of the Weyl determinant $\Delta_{Sp(2n)}(X)$ of an element
$X\in\mathfrak{sp}(2n)$ with eigenvalues $\mu_1,\dots,\mu_n$ (see
(\ref{pXdef}) and Table~\ref{Gtable}), as appears in the expression
(\ref{HCformula}). Furthermore, the Weyl group $W_{Sp(2n)}$ is
composed of permutations of eigenvalues $x_i\mapsto x_{\pi(i)}$,
$i=1,\dots n$, $\pi\in S_n$, along with reflections
$x_i\mapsto-x_i$. After the appropriate radial integration, the sum over
$W_{Sp(2n)}$ can thereby produce the required Pfaffian in the
numerator of the expression (\ref{Z12nRMT}), which in the random
matrix theory calculation arises naturally from quaternion
determinants generated in the skew-orthogonal polynomial formalism. In
the next subsection shall explicitly
work out the $\sigma$-model partition functions in the case of equal
quark masses and thus provide further evidence that the direct field
theory calculation indeed does reproduce this Pfaffian expression.

\subsubsection*{\mbf{N_f=2n+1}}

The restricted root lattice of the symmetric space
(\ref{Goldstone1odd}) is again the root space $BC_n$, with
multiplicities $m_{\vec e_i\pm\vec e_j}=4$, $m_{2\vec e_i}=3$ as
above, and in addition $m_{\vec e_i}=4$~\cite{Helg1,CM1}. The coset
space measure (\ref{Jacradial}) in this case is thus given by
\bea
\dd\nu_1(U)&=&[\dd V]~\prod_{i=1}^n\dd r_i~\sin^32r_i\,\sin^4r_i~
\prod_{i<j}\sin^4(r_i-r_j)\,\sin^4(r_i+r_j)\nonumber\\ &=&
\frac{[\dd V]}{2^{n\,(2n+1)}}~
\prod_{i=1}^n\dd\lambda_i~\left(1-\lambda_i^2\right)~
\left(1-\lambda_i\right)^2~\Delta(\lambda_1,\dots,\lambda_n)^4 \ .
\label{dnu1Uodd}\eea
With the definitions (\ref{rhodef}) and (\ref{xidef}), the radial
element
\beq
R=\pmatrix{\sigma^2& \cr
  &\exp\left(\,\ii\,\sigma^1\otimes\xi\right)\cr}
\label{radial1odd}\eeq
satisfies the required condition (\ref{RGammaprop}) and also
$R^\top=R^{-1}=R^\dag$. The angular matrices $V\in USp(2n+2)\times
USp(2n)$ are decomposed as in (\ref{Vbeta2}) with $V_+\in USp(2n+2)$
and $V_-\in USp(2n)$, so that (\ref{VGammabeta1}) is obeyed. The
$\sigma$-model action can be worked out analogously to
(\ref{action2odd}) and (\ref{sigmaaction1}), giving
\bea
Z_1^{(2n+1)}(\mu_0,\mu_1,\dots,\mu_n)&=&\prod_{i=1}^n~\int\limits_{-1}^1
\dd\lambda_i~\left(1-\lambda_i^2\right)~\left(1-\lambda_i
\right)^2~\Delta(\lambda_1,\dots,\lambda_n)^4\nonu &&\times\,
\Omega_1^{(n+1)}(1,\lambda_1,\dots,\lambda_n;\mu_0,\mu_1,\dots,\mu_n)~
\Omega_1^{(n)}(\lambda_1,\dots,\lambda_n;\mu_1,\dots,\mu_n)\nonu && +~~
(-1)^{N_\Dirac}~\left\{\stackrel{\scriptstyle\mu_0\to-\mu_0}{
\scriptstyle\mu_i\to-\mu_i}\right\} \ .
\label{Z1symplodd}\eea
For $n=0$ this expression is understood as being equal to
$\Omega_1^{(1)}(1;\mu_0)+(-1)^{N_\Dirac}\,\Omega_1^{(1)}(1;-\mu_0)$.
In~\cite{NN1} an expression for the finite volume partition function
in this case is derived from random matrix theory in the case of a
vanishing unpaired quark mass $\mu_0=0$ (so that the matrix model
measure is positive and the partition function vanishes when $N_\Dirac$ is
odd). In the next subsection we derive a formula for the equal mass
limit of (\ref{Z1symplodd}) without this restriction.

\subsection{Sum Rules\label{SRSympl}}

\subsubsection*{\mbf{N_f=2n}}

As in the previous section, to work out the massive spectral sum rules
(\ref{sumrulegen}) in the even-flavoured case at $\beta=1$ it is
convenient to work directly with the
partition function (\ref{Z1sympleven}) for equal quark masses
$\mu_i=\mu~~\forall i=1,\dots,n$, which is determined entirely by the
$n$ radial Goldstone boson degrees of freedom and is simply given by
\beq
Z_1^{(2n)}(\mu,\dots,\mu)=\prod_{i=1}^n~\int\limits_{-1}^1
\dd\lambda_i~\left(1-\lambda_i^2\right)~\e^{4\mu\,\lambda_i}~
\Delta(\lambda_1,\dots,\lambda_n)^4 \ .
\label{Z1equal}\eeq
Because of the original fermion determinant required in
(\ref{QCD3partfnfull}) and the doubling of the flavour symmetry in
this case, we can anticipate the appearence of a Pfaffian expression
for the finite volume partition function, in contrast to the
determinant formula of the previous section. For this, we write the
power of the Vandermonde determinant in (\ref{Z1equal}) as the
determinant of a $2n\times2n$ matrix given by~\cite{Mehta1,NF1}
\beq
\Delta(\lambda_1,\dots,\lambda_n)^4=\det\left[\matrix{\lambda_i^{j-1}\cr
(j-1)\,\lambda_i^{j-2}}\right]_{\stackrel{\scriptstyle i=1,\dots,n}
{\scriptstyle j=1,\dots,2n}} \ .
\label{Van4detid}\eeq
Expanding the determinant into a sum over permutations $\hat w\in
S_{2n}$, and using the permutation symmetry of the integration measure
and domain in (\ref{Z1equal}) to appropriately rearrange rows, we may
bring the equal mass partition function into the form
\bea
Z_1^{(2n)}(\mu,\dots,\mu)&=&\sum_{\hat w\in S_{2n}}
{\rm sgn}\left(\hat w\right)~\prod_{i=1}^n~\int\limits_{-1}^1
\dd\lambda_i~\left(1-\lambda_i^2\right)~\e^{4\mu\,\lambda_i}
\nonu&&\times\,
\Bigl(\hat w(2i)-1\Bigr)~\lambda_i^{\hat w(2i)+\hat w(2i-1)-3} \ .
\label{Z1permsum}\eea
Let $S_{2n}^>$ denote the subgroup of $S_{2n}$ consisting of
``increasing'' permutations $\hat w$, i.e. those obeying $\hat
w(2i)>\hat w(2i-1)~~\forall i=1,\dots,n$. Then the expression
(\ref{Z1permsum}) can be written as
\bea
Z_1^{(2n)}(\mu,\dots,\mu)&=&\sum_{\hat w\in S_{2n}^>}
{\rm sgn}\left(\hat w\right)~\prod_{i=1}^n~\int\limits_{-1}^1
\dd\lambda_i~\left(1-\lambda_i^2\right)~\e^{4\mu\,\lambda_i}
\nonu&&\times\,\Bigl(\hat w(2i)-\hat w(2i-1)\Bigr)~
\lambda_i^{\hat w(2i)+\hat w(2i-1)-3} \ .
\label{Z1almostPfaff}\eea

The expression (\ref{Z1almostPfaff}) now has the standard form of a
Pfaffian~\cite{Mehta1} and we may thereby write it as
\beq
Z_1^{(2n)}(\mu,\dots,\mu)=\pfaff\,\mbf A_{2n}(4\mu) \ ,
\label{Z1PfaffA}\eeq
where $\mbf A_{2n}(x)$ is the $2n\times2n$ antisymmetric matrix with
elements
\bea
\mbf A_{2n}(x)_{ij}&=&(i-j)\,\int\limits_{-1}^1\dd\lambda~\left(1-\lambda^2
\right)~\lambda^{i+j-3}~\e^{x\,\lambda}\nn\\&=&
2(i-j)\,\Bigl(\sine^{(i+j-3)}(x)-\sine^{(i+j-1)}(x)\Bigr) \ .
\label{A2nxdef}\eea
These matrix elements can be alternatively expressed as
\beq
\mbf A_{2n}(x)_{ij}=2(i-j)\,\left(\frac\dd{\dd x}\right)^{i+j-3}
\int\limits_{-1}^1\dd s~s\,\int\limits_{-1}^s\dd\lambda~\e^{x\,
\lambda}=(i-j)\,\left(\frac\dd{\dd x}\right)^{i+j-3}
\frac{\sine'(x)}x \ .
\label{A2nalt}\eeq
The substitution of (\ref{A2nalt}) into (\ref{Z1PfaffA}) can be
straightforwardly shown to coincide with the equal mass limit of the
finite volume partition function (\ref{Z12nRMT}) obtained from the
orthogonal ensemble of random matrix theory.

Proceeding as in Section~\ref{SRFund} to rewrite derivatives of the
spectral sine-kernel in terms of the functions (\ref{Knxij}), the
equal mass partition function in the case of an even number of
pseudo-real quarks acquires the Pfaffian form
\beq
\begin{tabular}{|c|}\hline\\
$\displaystyle
Z_1^{(2n)}(\mu,\dots,\mu)=\pfaff\left[(i-j)\,\Bigl(\mbf K_{i+j-3}(4\mu)-
\mbf K_{i+j-1}(4\mu)\Bigr)\right]_{i,j=1,\dots,2n}
$\\\\\hline\end{tabular}
\label{Z1Pfafffinal}\eeq
It can be computed explicitly by using the Laplace expansion of the
Pfaffian~\cite{Mehta1}
\beq
\pfaff\,\mbf A=\sum_{j=1}^{2n}(-1)^{i+j}~(\mbf A)_{ij}~
\pfaff\,\mbf A_{ji,ji}
\label{PfaffLaplace}\eeq
for any fixed $i=1,\dots,2n$, where the Pfaffian minor $\mbf
A_{ji,ji}$ is obtained from the antisymmetric $2n\times2n$ matrix
$\mbf A$ by deleting both its $i$-th and $j$-th rows and columns. The
massive spectral sum rules can thus be extracted from (\ref{Knxij})
and (\ref{Z1Pfafffinal}) by the substitution $N_f\to2N_f$ in
(\ref{sumrulegen}) to account for the symplectic doubling of fermion
degrees of freedom in this case, and for the first few even numbers of
quark flavours we find
\bea
&\begin{tabular}{|l|}\hline\\
$\left.\Biggl\langle\,\sum\limits_{\zeta>0}\frac1{\zeta^2+\mu^2}\Biggr\rangle_1
{}~\right|_{N_f=2}=\frac1{4\mu\,\coth4\mu-1}-\frac3{16\mu^2}
$\\${~~~~}^{~~}_{~~}$\\
$\left.\Biggl\langle\,\sum\limits_{\zeta>0}\frac1{\zeta^2+\mu^2}
\Biggr\rangle_1~\right|_{N_f=4}=\frac1{16\mu^2}\,
\frac{36\mu\left(4\mu^2+5\right)\sinh8\mu-9\left(64\mu^2+5\right)
\sinh^24\mu-16\mu^2\left(256\mu^4+216\mu^2+27\right)}
{9\left(8\mu^2+1\right)\sinh^24\mu-36\mu\,\sinh8\mu+128\mu^6
\left(16\mu^2+9\right)}
$\\\\\hline\end{tabular}\nonu&
\label{sumruleexplbeta1}\eea
These results agree with (\ref{masslesssumrule}) in the limit
$\mu\to0$ of massless quarks.

\subsubsection*{\mbf{N_f=2n+1}}

The equal mass limit $\mu_i=\mu~~\forall i=0,1,\dots,n$ of the
partition function (\ref{Z1symplodd}) is given by
\bea
Z_1^{(2n+1)}(\mu,\dots,\mu)&=&\e^{2\mu}~\prod_{i=1}^n~\int\limits_{-1}^1
\dd\lambda_i~\left(1-\lambda_i^2\right)~\left(1-\lambda_i\right)^2~
\e^{4\mu\,\lambda_i}~\Delta(\lambda_1,\dots,\lambda_n)^4 \nonu &&
+~~(-1)^{N_\Dirac}~\bigl\{\mu\to-\mu\bigr\} \ .
\label{Z1equalodd}\eea
The calculation is identical to that of the even-flavoured case
above. The only difference is that the antisymmetric matrix
(\ref{A2nxdef}) now contains an additional measure factor $(1-\lambda)^2$,
which when expanded leads to the equal mass finite volume partition
function in the case of an odd number of symplectic fermions in the
Pfaffian form
\bea
&\begin{tabular}{|l|}\hline\\
$\scriptstyle Z_1^{(2n+1)}(\mu,\dots,\mu)=\e^{2\mu}~\pfaff\Bigl[(i-j)\,
\Bigl({\bf K}_{i+j-3}(4\mu)-2{\bf K}_{i+j-2}(4\mu)+2{\bf K}_{i+j}(4\mu)-
{\bf K}_{i+j+1}(4\mu)\Bigr)
\Bigr]_{i,j=1,\dots,2n}$\\$\scriptstyle\qquad\qquad
+\,(-1)^{N_\Dirac}~\e^{-2\mu}~\pfaff\Bigl[(i-j)\,
\Bigl({\bf K}_{i+j-3}(4\mu)+2{\bf K}_{i+j-2}(4\mu)-2{\bf K}_{i+j}(4\mu)-
{\bf K}_{i+j+1}(4\mu)\Bigr)\Bigr]_{i,j=1,\dots,2n}
$\\\\\hline\end{tabular} \nonu &
\label{Z1oddequalfinal}\eea
This expression also covers the case $N_f=1$ of a single fermion of
mass $\mu$, obtained by dropping the Pfaffians in
(\ref{Z1oddequalfinal}), which for an even number $N_\Dirac$ of Dirac
operator eigenvalues leads to the massive spectral sum rule
\beq
\begin{tabular}{|l|}\hline\\
$\displaystyle
\left.\Biggl\langle\,\sum\limits_{\zeta>0}\frac1{\zeta^2+\mu^2}\Biggr\rangle_1
{}~\right|_{N_f=1}=\frac{\tanh2\mu}{4\mu}
$\\\\\hline\end{tabular}
\label{sumruleexplbeta1odd}\eeq

Sum rules for higher numbers of flavours are also straightforward to
work out, but they become increasingly complicated and will not be
displayed here. In each case one finds that the massless spectral sum
rule (\ref{masslesssumrule}) for $\beta=1$ is indeed obeyed. The
equations (\ref{holeprob}), (\ref{smallprob}), (\ref{Z1Pfafffinal})
and (\ref{Z1oddequalfinal}) suffice to express to smallest eigenvalue
distribution of the QCD$_3$ Dirac operator for $SU(2)$ colour group in
a closed form for an arbitrary number of fermion flavours in the case
of equal quark masses.

\newsection{Adjoint Fermions \label{Real}}

Finally, we consider the case of quark spinors which transform in the
adjoint representation of the $SU(N_c)$ colour group for $N_c\geq2$,
which is real. As in the previous section, the adjoint representation
of the Euclidean Dirac operator $\ii\,\Dirac$ uniquely defines an
anti-unitary operator $\mathfrak{C}_0$ with
$[\mathfrak{C}_0,\ii\,\Dirac]=0$, but now
$(\mathfrak{C}_0)^2=-\id$~\cite{HV1,Magnea2}. The operator
$\mathfrak{C}_0$ is given as a combination of charge conjugation
$C=\ii\,\sigma^2$ and complex conjugation. In this case the Dirac
operator can be expressed in terms of real quaternions~\cite{Magnea2},
and the condition $\mathfrak{C}_0|\psi_k\rangle=|\psi_k\rangle$
defines a basis $|\psi_k\rangle$ of Majorana spinors of the second
kind. The generic unitary $U(N_f)$ flavour symmetry thereby truncates
to its real orthogonal subgroup $O(N_f)$. In matrix model language
this corresponds to the symplectic ensemble $\beta=4$. The mass and
parity matrices are given exactly as in (\ref{MGamma2}), and so the
Goldstone manifolds are
\beq
\mathcal{G}_4(2n)&=&O(2n)\,/\,O(n)\times O(n) \ , \label{Goldstone4}\\
\mathcal{G}_4(2n+1)&=&O(2n+1)\,/\,O(n+1)\times O(n)
\label{Goldstone4odd}\eeq
of real dimensions $\dim\mathcal{G}_4(2n)=n^2$ and
$\dim\mathcal{G}_4(2n+1)=n\,(n+1)$. As expected, the number
of Goldstone bosons here is exactly half that of Section~\ref{FF}.

\subsection{Partition Functions\label{PFAd}}

\subsubsection*{$\mbf{N_f=2}$}

The ``degenerate'' case $N_f=2$ corresponding to two-flavour component
adjoint fermions is very special because it represents flavour
symmetry breaking from $O(2)$ to a {\it finite} subgroup
$\zed_2\times\zed_2$. The coset space (\ref{Goldstone4}) in this case
is the orbifold
\beq
\mathcal{G}_4(2)~=~O(2)\,/\,\zed_2\times\zed_2 \ .
\label{G41orbifold}\eeq
The quotient by one of the $\zed_2$ factors in (\ref{G41orbifold}) can
be used to eliminate one of the connected components of the $O(2)$
group, thereby realizing the orbifold explicitly as a one-dimensional
manifold diffeomorphic to an open interval
\beq
\mathcal{G}_4(2)~=~SO(2)\,/\,\zed_2~\cong~(0,\pi) \ .
\label{Goldstoneinterval}\eeq
After flavour symmetry breaking, there is thus a single massless
degree of freedom described in the mesoscopic scaling regime by a
real scalar field $\theta\in(0,\pi)$.

To evaluate the corresponding finite volume partition function
(\ref{ZMgen}), we parametrize the elements $U\in SO(2)$ appearing in
(\ref{Goldstoneinterval}) as
\beq
U=\pmatrix{\cos\theta&-\sin\theta\cr\sin\theta&\cos\theta}
\label{USO2par}\eeq
and work out the effective $\sigma$-model action to be
\beq
{\rm Re}\,\Tr\left(M\,U\,\Gamma\,U^\top\,\right)=2\mu\,\cos2\theta \ .
\label{sigmaactiondeg}\eeq
With the normalized Haar measure $[\dd U]=\dd\theta/\pi$ on $SO(2)$, the
partition function (\ref{ZMgen}) in the two flavour case is thereby
found to be
\beq
\begin{tabular}{|c|}\hline\\
$\displaystyle
Z_4^{(2)}(\mu)=I_0(2\mu)
$\\\\\hline\end{tabular}
\label{2flavourpart}\eeq
where generally
\beq
I_m(x)=\int\limits_{-\pi}^\pi\frac{\dd\phi}{2\pi}~\e^{x\,\cos\phi+
\ii\,m\,\phi}
\label{BesselIdef}\eeq
is the modified Bessel function of the first kind of integer order
$m$.

\subsubsection*{$\mbf{N_f=2n}$}

Let us now turn to the case of generic even integer values $N_f=2n$
with $n>1$. The restricted root lattice of the symmetric space
(\ref{Goldstone4}) is the root space $D_n$ of the special orthogonal
group $SO(2n)$ (see Table~\ref{Gtable}), with root multiplicities
$m_{\vec e_i\pm\vec e_j}=1$ for $i,j=1,\dots,n$, $i<j$~\cite{Helg1,CM1}. The
corresponding radial integration measure (\ref{Jacradial}) on
$\mathcal{G}_4(2n)$ is thus
\beq
\dd\nu_4(U)=[\dd V]~\prod_{i=1}^n\dd r_i~\prod_{i<j}\,\Bigl|
\sin(r_i-r_j)\,\sin(r_i+r_j)\Bigr| \ .
\label{radial4}\eeq
Similarly to the $n=1$ case, the quotient in (\ref{Goldstone4}) can be
used to eliminate the connected components of the orthogonal groups,
at the expense of presenting the coset space as the orbifold
\beq
\mathcal{G}_4(2n)~=~SO(2n)\,/\,SO(n)\times SO(n)\times\zed_2 \ .
\label{G4norbifold}\eeq
In this case, it is more convenient to parametrize the radial
directions of this orbifold in terms of the angle
variables $\theta_i=2r_i\in[0,\pi]$, and using (\ref{trigid}) we may
write (\ref{radial4}) as
\beq
\dd\nu_4(U)=\frac{[\dd V]}{2^{\frac12\,n\,(n+1)}}~
\prod_{i=1}^n\dd\theta_i~\Bigl|\Delta(
\cos\theta_1,\dots,\cos\theta_n)\Bigr| \ .
\label{dnu4Ufinal}\eeq
To ensure positivity of this measure when the absolute value signs are
dropped, we will restrict the integration region for the $\theta_i$'s
to the domain
\beq
\mathcal{D}^\theta_n=\Bigl\{(\theta_1,\dots,\theta_n)~\Bigm|~
0\leq\theta_1\leq\theta_2\leq\dots\leq\theta_n\leq\pi\Bigr\} \ .
\label{thetadomaindef}\eeq
This can always be achieved without complications for the integration of
symmetric functions.

The particular details of the radial decomposition here are identical
to those of Section~\ref{PFFund}, and we thereby arrive at the
non-linear $\sigma$-model action
\bea
{\rm Re}\,\Tr\left(M\,U\,\Gamma\,U^\top\,\right)&=&\Tr\left[
\pmatrix{\scriptstyle\mu_1& & \cr
 &\scriptstyle\ddots& \cr & &\scriptstyle\mu_n}\,V^{~}_+\,
\pmatrix{\scriptstyle\cos\theta_1& & \cr
 &\scriptstyle\ddots& \cr & &\scriptstyle\cos\theta_n}\,
V_+^{-1}\right]\nonu&&+\,\Tr\left[\pmatrix{\scriptstyle\mu_1& & \cr
 &\scriptstyle\ddots& \cr & &\scriptstyle\mu_n}\,V^{~}_-\,
\pmatrix{\scriptstyle\cos\theta_1& & \cr
 &\scriptstyle\ddots& \cr & &\scriptstyle\cos\theta_n}\,
V_-^{-1}\right]
\label{sigmaaction4}\eeq
where now $V_\pm\in O(n)$. The finite volume partition function is
thus given by
\beq
Z_4^{(2n)}(\mu_1,\dots,\mu_n)=\int\limits_{\mathcal{D}_n^\theta}\,
\prod_{i=1}^n\dd\theta_i~\Delta(\cos\theta_1,\dots,\cos\theta_n)~
\Omega_4^{(n)}(\cos\theta_1,\dots,\cos\theta_n;\mu_1,\dots,\mu_n)^2
\label{Z42ngen}\eeq
where
\bea
&&\Omega_4^{(n)}(x_1,\dots,x_n;\mu_1,\dots,\mu_n)\nonu&&\qquad\qquad~=~
\int\limits_{SO(n)}[\dd V]~\exp\,\Tr\left[\pmatrix{\scriptstyle\mu_1& & \cr
 &\scriptstyle\ddots& \cr & &\scriptstyle\mu_n}\,V\,
\pmatrix{\scriptstyle x_1& & \cr
 &\scriptstyle\ddots& \cr & &\scriptstyle x_n}\,
V^{-1}\right] \ .
\label{SOnintunknown}\eea
As in the previous section, the integral (\ref{SOnintunknown}) cannot
be evaluated using the Harish-Chandra formula (\ref{HCformula})
because the Lie algebra $\mathfrak{so}(n)$ consists of antisymmetric
$n\times n$ matrices. The first two members of this sequence of
angular integrals are easily evaluated explicitly to be
\bea
\Omega_4^{(1)}(x;\mu)&=&\e^{\mu\,x} \ , \nonu
\Omega_4^{(2)}(x_1,x_2;\mu_1,\mu_2)&=&2~
\e^{\frac12\,(\mu_1+\mu_2)\,(x_1+x_2)}~I_0\left(\mbox{$\frac12$}\,
(\mu_1-\mu_2)\,(x_1-x_2)\right) \ .
\label{Omega4expls}\eea
For $n=3$ the integral (\ref{SOnintunknown}) can be computed by means
of the Harish-Chandra formula for $SU(2)\cong SO(3)$.

In~\cite{NN1} the finite volume partition function in this instance is
computed using a skew-orthogonal polynomial formalism in the
symplectic ensemble of random matrix theory. For $n=2\ell$ with
$\ell\geq1$ it is necessary to assume that the mass eigenvalues
$\mu_i$ appear in degenerate pairs, i.e. $\mu_{2a-1}=\mu_{2a}$ for
each $a=1,\dots,\ell$, in order to ensure positivity of the matrix
model measure. In this case a Pfaffian expression for the partition function
is obtained in terms of integrals of the spectral sine kernel
(\ref{sinexdef}) (rather than derivatives as in the previous
section), or equivalently in terms of exponential integral functions
${\rm Ei}(x)$. In general, this restriction does not aid the field theory
calculation, which should thereby capture the case of generic mass
eigenvalue configurations. In the next subsection we will evaluate
(\ref{Z42ngen}) in the equal mass limit and find a striking contrast
between the field theory and random matrix theory expressions. As in
(\ref{2flavourpart}), the field theory calculation will naturally
involve modified Bessel functions of the first kind (along with sine
kernel derivatives), which is reminescent of results for QCD in {\it
  four} spacetime dimensions~\cite{LS1,GW1}. This may be a consequence
of the known relations between the QCD$_3$ and QCD$_4$ partition
functions in the $\epsilon$-regime~\cite{ADDV1,AD1}. Comparison of
our expressions with those of~\cite{NN1} imply that some non-trivial
simplifying identities among these Bessel functions ought to hold,
but we have not been able to find any.

For $n=2\ell+1$, in addition to pairwise degeneracy, it is necessary to
assume in the random matrix theory approach that the single unpaired
quark mass vanishes, again for
positivity reasons. In the next subsection we derive the equal mass
limit of (\ref{Z42ngen}) without this restriction. From the field
theory perspective, we shall also encounter a distinction between the
cases $N_f=4\ell$ and $N_f=4\ell+2$, and this distinction is very
natural. From Table~\ref{Gtable} we see that the Weyl group sum for
the group $SO(2\ell+1)$ is the same as that of the symplectic group
$Sp(2\ell)$. In contrast, the Weyl group of $SO(2\ell)$ contains the
usual permutation part $S_\ell$ and the factor $(\zed_2)^{\ell-1}$ is the
normal subgroup given by the possible assignments of an {\it even} number of
sign flips to the eigenvalues. These facts, along with observations
similar to those spelled out in the previous section, support the
precise matching between the field theory and random matrix theory
results in these instances.

\subsubsection*{$\mbf{N_f=2n+1}$}

The restricted root lattice of the Goldstone manifold
(\ref{Goldstone4odd}) is the root lattice $B_n$ of the special
orthogonal group $SO(2n+1)$ (see Table~\ref{Gtable}), with
multiplicities $m_{\vec e_i\pm\vec e_j}=m_{\vec e_i}=1$ for
$i,j=1,\dots,n$, $i<j$~\cite{Helg1,CM1}. The corresponding coset space
measure (\ref{Jacradial}) is thus

\vbox{\bea
\dd\nu_4(U)&=&[\dd V]~\prod_{i=1}^n\dd r_i~\sin r_i~\prod_{i<j}\,\Bigl|
\sin(r_i-r_j)\,\sin(r_i+r_j)\Bigr| \nonu &=&[\dd V]~\prod_{i=1}^n
\dd\kappa_i~\left|\Delta\left(\kappa_1^2,\dots,\kappa_n^2\right)\right| \ ,
\label{dnu4Uodd}\eea}\noindent
where we have defined $\kappa_i=\cos r_i\in[0,1]$ and used (\ref{trigid})
along with the double angle identity. By introducing the domain
\beq
\mathcal{D}^\kappa_n=\Bigl\{(\kappa_1,\dots,\kappa_n)~\Bigm|~
0\leq\kappa_1\leq\kappa_2\leq\dots\leq\kappa_n\leq1\Bigr\}
\label{xidomaindef}\eeq
and proceeding exactly as in Section~\ref{PFFund}, the finite volume
partition function may then be written as
\bea
Z_4^{(2n+1)}(\mu_0,\mu_1,\dots,\mu_n)&=&\int\limits_{\mathcal{D}_n^\kappa}\,
\prod_{i=1}^n\dd\kappa_i~\Delta\left(\kappa_1^2,\dots,\kappa_n^2\right)
{}~\Omega_4^{(n)}\left(2\kappa_1^2-1,
\dots,2\kappa_n^2-1;\mu_1,\dots,\mu_n\right) \nonu &&
\times~\Omega_4^{(n+1)}\left(1,2\kappa_1^2-1,\dots,2\kappa_n^2-1;
\mu_0,\mu_1,\dots,\mu_n\right) \nonu &&
+~~(-1)^{N_\Dirac}~\left\{\stackrel{\scriptstyle\mu_0\to-\mu_0}
{\scriptstyle\mu_i\to-\mu_i}\right\}
\label{Z42n1gen}\eea
with the same conventions for the $n=0$ case as in (\ref{Z1symplodd}). The
odd-flavoured cases are not covered by the random matrix theory
analysis of~\cite{NN1}. In the next subsection we will compute
(\ref{Z42n1gen}) explicitly in the case of equal quark masses. We will
find that, in contrast to the previous two classes, for adjoint
fermions they involve fundamentally different functional structures
than those of the even-flavoured cases. In this case the partition
function is expressed in terms of a combination of error functions and
Hermite polynomials, and appears to be more akin to the random matrix
theory predictions~\cite{NN1,Mehta1}. These functional forms are
similiar to those which arise in partially-quenched effective field
theories of QCD$_3$~\cite{me}.

\subsection{Sum Rules\label{SRAd}}

To compute the massive spectral sum rules (\ref{sumrulegen}), we
proceed as previously to compute directly the finite volume partition
functions (\ref{ZMgen}) in the limit of equal quark masses for the
various congruence classes modulo~$4$ of the flavour number $N_f$.

\subsubsection*{$\mbf{N_f=2}$}

For $N_f=2$, we substitute (\ref{2flavourpart}) into (\ref{sumrulegen})
to compute
\beq
\begin{tabular}{|l|}\hline\\
$\displaystyle\left.\left\langle\,\sum_{\zeta>0}
\frac1{\zeta^2+\mu^2}\right\rangle_4
{}~\right|_{N_f=2}=\frac1{2\mu}\,\frac{I_1(2\mu)}{I_0(2\mu)}
$\\\\\hline\end{tabular}
\label{sumruleNf2beta1}\eeq
Once again this agrees with (\ref{masslesssumrule}) in the limit
$\mu\to0$.

\subsubsection*{$\mbf{N_f=4\ell}$}

For $N_f=2n$ with $n>1$ the partition function (\ref{Z42ngen}) for
equal quark masses is given by
\beq
Z_4^{(2n)}(\mu,\dots,\mu)=\int\limits_{\mathcal{D}^\theta_n}\,\prod_{i=1}^n
\dd\theta_i~\e^{2\mu\,\cos\theta_i}~\Delta(\cos\theta_1,
\dots,\cos\theta_n) \ .
\label{Z4neqmassgen}\eeq
Again, since the functional integration in the original QCD$_3$
partition function is over {\it real} Majorana fermion fields, thereby
producing a square-root fermion determinant in (\ref{QCD3partfnfull}),
we can anticipate a Pfaffian expression for (\ref{Z4neqmassgen}). With
this in mind, we will proceed to evaluate the finite volume partition
function using the method of integration over alternate
variables~\cite{Mehta1,NF1}, whose details again depend crucially on
the even/odd parity of the flavour rank $n$.

For $n=2\ell$ with $\ell\geq1$ we can insert the determinant
representation (\ref{Vandermondedet}) to write
\bea
Z_4^{(4\ell)}(\mu,\dots,\mu)&=&\int\limits_{\mathcal{D}^\theta_{2\ell}}
\,\prod_{a=1}^\ell\dd\theta_{2a}~\e^{2\mu\,\cos\theta_{2a}}~\int
\limits_{\theta_{2a-2}}^{\theta_{2a}}\dd\theta_{2a-1}~\e^{2\mu\,
\cos\theta_{2a-1}}~\det\left[\cos^{j-1}\theta_i
\right]_{i,j=1,\dots,2\ell}\nonu&&
\label{Z42elldetrep}\eea
where we have defined $\theta_0=0$. We formally insert the
integrals into the determinant in (\ref{Z42elldetrep}) by using its
multilinearity to write
\beq
Z_4^{(4\ell)}(\mu,\dots,\mu)=\int\limits_{\mathcal{D}^\theta_{2\ell}}
\,\prod_{a=1}^\ell\dd\theta_{2a}~\e^{2\mu\,\cos\theta_{2a}}~
\det\left[\matrix{\cos^{j-1}\theta_{2b}\cr
\,\int\limits_{\theta_{2b-2}}^{\theta_{2b}}
\dd\theta~\e^{2\mu\,\cos\theta}~\cos^{j-1}\theta}
\right]_{\stackrel{\scriptstyle
b=1,\dots,\ell}{\scriptstyle j=1,\dots,2\ell}} \ .
\label{Z42ellmultilin}\eeq
Since the integrand of (\ref{Z42ellmultilin}) is a symmetric function
of the integration variables, we may extend the lower limit of
integration of each variable inside the determinant down to~$0$. This
symmetry also allows us to extend the range of each of the remaining
integration variables to $[0,\pi]$. After rearranging rows in the
determinant and expanding it as a sum over permutations, we arrive at
the expression
\bea
Z_4^{(4\ell)}(\mu,\dots,\mu)&=&\sum_{\hat w\in S_{2\ell}}
{\rm sgn}\left(\hat w\right)~\prod_{a=1}^\ell~\int\limits_0^\pi
\dd\theta_{2a}~\e^{2\mu\,\cos\theta_{2a}}~
\cos^{\hat w(2a)-1}\theta_{2a}\nonu&&
\times\,\int\limits_0^{\theta_{2a}}\dd\theta~
\e^{2\mu\,\cos\theta}~\cos^{\hat w(2a-1)-1}\theta
\label{Z42elldetexp}\eea
which may be written as a sum over the subgroup $S_{2\ell}^>\subset
S_{2\ell}^{~}$ of increasing permutations given by
\bea
Z_4^{(4\ell)}(\mu,\dots,\mu)&=&\sum_{\hat w\in S^>_{2\ell}}
{\rm sgn}\left(\hat w\right)~\prod_{a=1}^\ell~\int\limits_0^\pi
\dd\theta_{2a}~\e^{2\mu\,\cos\theta_{2a}}~
\int\limits_0^{\theta_{2a}}\dd\theta~\e^{2\mu\,\cos\theta}\nonu&&
\times\,\left(\cos^{\hat w(2a)-1}\theta_{2a}~
\cos^{\hat w(2a-1)-1}\theta-\cos^{\hat w(2a-1)-1}\theta_{2a}~
\cos^{\hat w(2a)-1}\theta\right) \ . \nonu&&
\label{Z42elldetincr}\eea

As before, the expression (\ref{Z42elldetincr}) now has the standard
form of a Pfaffian and we may write it as
\beq
Z_4^{(4\ell)}(\mu,\dots,\mu)=\pfaff\,\mbf B_{2\ell}(2\mu) \ ,
\label{Z4PfaffB}\eeq
where $\mbf B_{2\ell}(x)$ is the $2\ell\times2\ell$ antisymmetric
matrix with elements
\beq
\mbf B_{2\ell}(x)_{ij}=\int\limits_0^\pi\dd\theta~\e^{x\,\cos\theta}~
\cos^{i-1}\theta~\int\limits_0^\pi\dd\theta'~
{\rm sgn}\left(\theta-\theta'\,\right)~\e^{x\,\cos\theta'}~
\cos^{j-1}\theta' \ .
\label{B2ellxdef}\eeq
The integrals in (\ref{B2ellxdef}) can be evaluated explicitly by
using the Fourier series representation of the sign function
\beq
{\rm sgn}(\phi)=\frac4\pi\,\sum_{k=0}^\infty\frac{\sin(2k+1)\phi}
{2k+1}
\label{sgnphiFourier}\eeq
for $\phi\in[-\pi,\pi]$, and defining the sequence of functions
\beq
\sinec_{2k+1}(x)=\int\limits_0^\pi\dd\theta~\e^{x\,\cos\theta+
(2k+1)\,\ii\,\theta}
\label{sinecmxdef}\eeq
for $k\in\nat_0$. We may then express the matrix elements
(\ref{B2ellxdef}) as series in derivatives of the functions
(\ref{sinecmxdef}) given by
\beq
\mbf B_{2\ell}(x)_{ij}=\frac\ii{2\pi}\,\sum_{k=0}^\infty
\frac{\sinec^{(i-1)}_{2k+1}(x)\,\sinec^{(j-1)}_{2k+1}(x)^*-
\sinec^{(j-1)}_{2k+1}(x)\,\sinec^{(i-1)}_{2k+1}(x)^*}{2k+1} \ .
\label{B2ellCseries}\eeq

Let us now explicitly evaluate the derivatives of the functions
(\ref{sinecmxdef}). The real parts can be written in terms of the
modified Bessel functions (\ref{BesselIdef}) as
\beq
{\rm Re}~\sinec_{2k+1}^{(i-1)}(x)=\pi\,I_{2k+1}^{(i-1)}(x) \ .
\label{Resinec}\eeq
For the imaginary parts, we use the expansion
\bea
\sin(2k+1)\theta&=&{\rm Im}\,(\cos\theta+\ii\,\sin\theta)^{2k+1}
\nonumber\\&=&\sin\theta\,\sum_{l=0}^k{2k+1\choose2l+1}\,\sum_{a=0}^l
(-1)^a\,{l\choose a}\,\cos^{2k-2a}\theta
\label{sineexpand}\eea
along with the change of integration variable
$\lambda=\cos\theta\in[-1,1]$ in (\ref{sinecmxdef}) to evaluate them
in terms of the sine kernel derivatives (\ref{Knxint}) as
\beq
{\rm Im}~\sinec_{2k+1}^{(i-1)}(x)=
2\,\sum_{l=0}^k{2k+1\choose 2l+1}\,\sum_{a=0}^l(-1)^a\,
{l\choose a}\,\sine^{(2k-2a+i-1)}(x) \ .
\label{sinecckidef}\eeq
The sums over $l$ and $a$ in (\ref{sinecckidef}) can be interchanged,
after which the sum over $p=l-a$ for fixed $a$ can be evaluated
explicitly by using the combinatorial identity
\beq
\sum_{p=0}^{k-a}{2k+1\choose2p+2a+1}\,{p+a\choose a}=
4^{k-a}\,{2k-a\choose a}
\label{combinid}\eeq
that is straightforwardly proven by induction.
Substituting in the explicit functional forms (\ref{Knxij}) for the
sine kernel derivatives, and plugging (\ref{Resinec}) and
(\ref{sinecckidef}) into (\ref{B2ellCseries}) then enables us to write
the equal mass partition function (\ref{Z4PfaffB}) explicitly as
\bea
&\begin{tabular}{|l|}\hline\\
$\displaystyle Z_4^{(4\ell)}(\mu,\dots,\mu)=\pfaff\left[
\,\sum_{k=0}^\infty\frac{I_{2k+1}^{(i-1)}(2\mu)\,\sinecc_{2k;j-1}
(2\mu)-I_{2k+1}^{(j-1)}(2\mu)\,\sinecc_{2k;i-1}
(2\mu)}{2k+1}\right]_{i,j=1,\dots,2\ell}
$\\\\\hline\end{tabular}\nonu&
\label{Z4Pfafffinal}\eea
where
\bea
\sinecc_{k;i}(x)&=&\frac12\,\sum_{b=0}^{k}~
\sum_{a=0}^{\lfloor\frac{k+i-b}2\rfloor}\,
\frac{(-1)^a\,2^{k-2a}\,(k-2a+i)!}{(k-2a-b+i)!}\,
{k-a\choose a}\,\left(-\frac1x\right)^{b+1}\nonu&&
\times\,\left[\Bigl(1-(-1)^b\Bigr)\,
\Bigl(\sinh x-\cosh x\Bigr)-2\sinh x\right] \ .
\label{sinecckiexpl}\eea

{}From (\ref{Z4Pfafffinal}) one may now extract the massive spectral sum
rules (\ref{sumrulegen}) by using the derivative identities
\beq
\frac{\dd I_m(x)}{\dd x}=I_{m+1}(x)+\frac mx\,I_m(x) \ , ~~
\frac{\dd\sinecc_{k;i}(x)}{\dd x}=\sinecc_{k;i+1}(x) \ .
\label{derivids}\eeq
For instance, for four flavours of adjoint fermions one finds
\beq
\begin{tabular}{|l|}\hline\\
$\left.\Biggl\langle\,\sum\limits_{\zeta>0}\frac1{\zeta^2+\mu^2}
\Biggr\rangle_4~\right|_{N_f=4}
=\frac{\sum\limits_{k=0}^\infty
\frac{I_{2k+1}(2\mu)\,\left(\sineccs_{k;2}(2\mu)-\frac{4k^2-1}
{4\mu^2}\,\sineccs_{k;0}(2\mu)\right)-\sineccs_{k;0}(2\mu)\,
\left(I_{2k+3}(2\mu)+\frac{4k+3}{2\mu}\,I_{2k+2}(2\mu)\right)}
{2k+1}}{4\mu\,\sum\limits_{k=0}^\infty\frac{I_{2k+1}(2\mu)\,
\left(\sineccs_{k;1}(2\mu)-\frac{2k+1}{2\mu}\,\sineccs_{k;0}
(2\mu)\right)-I_{2k+2}(2\mu)\,\sineccs_{k;0}(2\mu)}{2k+1}}
$\\\\\hline\end{tabular}\nonu
\label{beta4flav4sumrule}\eeq
A numerical evaluation of (\ref{beta4flav4sumrule}) in the limit
$\mu\to0$ establishes precise agreement with (\ref{masslesssumrule})
in this case.

\subsubsection*{$\mbf{N_f=4\ell+2}$}

Let us now consider the case $N_f=2n$ with generic odd flavour rank
$n=2\ell+1$, $\ell\geq1$. The calculation proceeds analogously to that
of the even rank case, except that now one needs to keep careful track
of the single unpaired integration. Starting from (\ref{Z4neqmassgen})
we derive the determinant formula

\vbox{\bea
Z_4^{(4\ell+2)}(\mu,\dots,\mu)&=&\int\limits_{\mathcal{D}^\theta_{2\ell+1}}
\,\prod_{a=1}^\ell\dd\theta_{2a}~\e^{2\mu\,\cos\theta_{2a}}~\int
\limits_{\theta_{2a-2}}^{\theta_{2a}}\dd\theta_{2a-1}~\e^{2\mu\,
\cos\theta_{2a-1}}\nonu &&\times\,\det\left[\cos^{j-1}\theta_i
\right]_{i,j=1,\dots,2\ell+1}~\int\limits_{\theta_{2\ell}}^\pi
\dd\theta_{2\ell+1}~\e^{2\mu\,\cos\theta_{2\ell+1}}\nonumber\\
&& {~~~~}^{~~}_{~~} \nn\\ &=&
\int\limits_{\mathcal{D}^\theta_{2\ell+1}}
\,\prod_{a=1}^\ell\dd\theta_{2a}~\e^{2\mu\,\cos\theta_{2a}}~
\det\left[\matrix{\cos^{j-1}\theta_{2b}\cr
\int\limits_{\theta_{2b-2}}^{\theta_{2b}}
\dd\theta~\e^{2\mu\,\cos\theta}~\cos^{j-1}\theta\cr
\,\int\limits_{\theta_{2\ell}}^\pi\dd\phi~\e^{2\mu\,\cos\phi}~
\cos^{j-1}\phi}\right]_{\stackrel{\scriptstyle
b=1,\dots,\ell}{\scriptstyle j=1,\dots,2\ell+1}} \ . \nonu&&
\label{Z42ell1detrep}\eea}
\noindent
Using symmetry of the integrand in (\ref{Z42ell1detrep}) to extend the
integration ranges as before, expansion of
the determinant into a sum over permutations yields
\bea
Z_4^{(4\ell+2)}(\mu,\dots,\mu)&=&\sum_{\hat w\in S_{2\ell+1}}
{\rm sgn}\left(\hat w\right)~\prod_{a=1}^\ell~\int\limits_0^\pi
\dd\theta_{2a}~\e^{2\mu\,\cos\theta_{2a}}~
\cos^{\hat w(2a)-1}\theta_{2a}\nonu&&
\times\,\int\limits_0^{\theta_{2a}}\dd\theta~
\e^{2\mu\,\cos\theta}~\cos^{\hat w(2a-1)-1}\theta~
\int\limits_0^\pi\dd\phi~\e^{2\mu\,\cos\phi}~
\cos^{\hat w(2\ell+1)-1}\phi\nonumber\\
&& {~~~~}^{~~}_{~~} \nn\\ &=&\sum_{\hat w\in S^>_{2\ell+1}}
{\rm sgn}\left(\hat w\right)~\prod_{a=1}^\ell~\int\limits_0^\pi
\dd\theta_{2a}~\e^{2\mu\,\cos\theta_{2a}}~
\int\limits_0^{\theta_{2a}}\dd\theta~\e^{2\mu\,\cos\theta}\nonu&&
\times\,\left(\cos^{\hat w(2a)-1}\theta_{2a}~
\cos^{\hat w(2a-1)-1}\theta-\cos^{\hat w(2a-1)-1}\theta_{2a}~
\cos^{\hat w(2a)-1}\theta\right)\nonu&&\times\,
\int\limits_0^\pi\dd\phi~\e^{2\mu\,\cos\phi}~
\cos^{\hat w(2\ell+1)-1}\phi
\label{Z42ell1detexp}\eea
which again has the standard form of a Pfaffian of dimension
$(2\ell+2)\times(2\ell+2)$. The last integral in (\ref{Z42ell1detexp})
can be expressed in terms of derivatives of the modified Bessel
function~$I_0(2\mu)$. With the same functions (\ref{sinecckiexpl}), we
thereby find the equal mass finite volume partition function in the
explicit form
\bea
&\begin{tabular}{|l|}\hline\\
$Z_4^{(4\ell+2)}(\mu,\dots,\mu)$\\$~~~~~~~~~~~~~~~=\pfaff\left[
\matrix{\,\sum\limits_{k=0}^\infty
\frac{I_{2k+1}^{(i-1)}(2\mu)\,\sineccs_{k;j-1}
(2\mu)-I_{2k+1}^{(j-1)}(2\mu)\,\sineccs_{k;i-1}
(2\mu)}{2k+1}&\scriptstyle
I_0^{(i-1)}(2\mu)\cr\scriptstyle-I_0^{(j-1)}(2\mu)&\scriptstyle0}
\right]_{i,j=1,\dots,2\ell+1}
$\\\\\hline\end{tabular}\nonu&
\label{Z41Pfafffinal}\eea
The (rather cumbersome) sum rules may now be extracted from
(\ref{Z41Pfafffinal}) exactly as described above for the even rank
case.

\subsubsection*{\mbf{N_f=4\ell+1}}

For odd numbers of flavours $N_f=2n+1$ the equal mass limit of the
finite volume partition function (\ref{Z42n1gen}) is given by
\beq
Z_4^{(2n+1)}(\mu,\dots,\mu)=\e^{-(2n-1)\,\mu}~\int
\limits_{\mathcal{D}_n^\kappa}\,\prod_{i=1}^n\dd\kappa_i~\e^{4\mu\,\kappa_i^2}~
\Delta\left(\kappa_1^2,\dots,\kappa_n^2\right)~~+~~(-1)^{N_\Dirac}~\bigl\{
\mu\to-\mu\bigr\} \ . \nonu
\label{Z4equaloddgen}\eeq
The calculation of the integrals in (\ref{Z4equaloddgen}) is identical
to that of (\ref{Z4neqmassgen}), with the obvious modification of the
integration domain and the replacements $\cos\theta^{~}_i\to2\kappa_i^2$
everywhere in the previous formulas. For $n=2\ell$, one may in this
way bring (\ref{Z4equaloddgen}) into the Pfaffian form
\beq
Z_4^{(4\ell+1)}(\mu,\dots,\mu)=\e^{-(4\ell-1)\,\mu}~\pfaff\,\mbf
E_{2\ell}(4\mu)+(-1)^{N_{\Dirac}}~\e^{(4\ell-1)\,\mu}~\pfaff\,\mbf
E_{2\ell}(-4\mu)
\label{Z42elloddgen}\eeq
where $\mbf E_{2\ell}(x)$ is the $2\ell\times2\ell$ antisymmetric matrix
defined by the elements
\beq
\mbf E_{2\ell}(x)_{ij}=\int\limits_0^1\dd\kappa~\e^{x\,\kappa^2}~
\left(\kappa^2\right)^{i-1}~\int\limits_0^1\dd\kappa'~{\rm
sgn}\left(\kappa-\kappa'\,\right)~\e^{x\,\kappa^{\prime\,2}}~
\left(\kappa^{\prime\,2}\right)^{j-1} \ .
\label{E2elldef}\eeq
In this case it is more convenient to use the Fourier integral
representation of the sign function given by
\beq
{\rm sgn}(y)=\frac1\pi~\int\limits_{-\infty}^\infty\dd q~\frac{\sin
  q\,y}{q+\ii\,\varepsilon}
\label{sgnFourierint}\eeq
for $y\in\real$, where the infinitesimal parameter $\varepsilon\to0^+$
regulates the ambiguity in ${\rm sgn}(y)$ at $y=0$. Then the matrix
elements (\ref{E2elldef}) can be written in terms of the functions
\beq
\Xi_q(x)=\int\limits_0^1\dd\kappa~\e^{x\,\kappa^2+\ii\,q\,\kappa}
\label{Xiqxdef}\eeq
which may be integrated explicitly in terms of error functions.

In this we may write the finite volume partition function
(\ref{Z42elloddgen}) explicitly as
\bea
&\begin{tabular}{|l|}\hline\\
$Z_4^{(4\ell+1)}(\mu,\dots,\mu)$\\$\qquad=\e^{-(4\ell-1)\,\mu}~
\pfaff\left[\frac\ii{4\pi}~\int\limits_{-\infty}^\infty\dd q~
\frac{\Xi_q^{(i-1)}(4\mu)\,\Xi_q^{(j-1)}(4\mu)^*-
\Xi_q^{(j-1)}(4\mu)\,\Xi_q^{(i-1)}(4\mu)^*}{q+\ii\,\varepsilon}
\right]_{i,j=1,\dots,2\ell}$\\$\qquad\quad+~~(-1)^{N_\Dirac}~
\bigl\{\mu\to-\mu\bigr\}
$\\\\\hline\end{tabular}\nonu&
\label{Z4ell1Pfafffinal}\eea
where
\beq
\Xi_q(x)=\mbox{$\frac1{2\,\ii}$}~\sqrt{\mbox{$\frac\pi x$}}~
\e^{\frac{q^2}{2x}}~\left[\erf\left(\mbox{$\frac q{2\,\sqrt
        x}$}\right)-\erf\left(\mbox{$\frac{q-2\,\ii\,x}{2\,\sqrt x}$}
\right)\right] \ .
\label{Xierf}\eeq
The matrix elements in (\ref{Z4ell1Pfafffinal}) can be made somewhat
more explicit by using the derivative identity
\beq
\frac{\dd^k\,\erf(x)}{\dd x^k}=(-1)^{k-1}~
\frac{2}{\sqrt\pi}~\e^{-x^2}~H_{k-1}(x) \ ,
\label{derfHermite}\eeq
where $H_m(x)$ are the Hermite polynomials. As before, this expression
includes the one-flavour case $N_f=1$ ($\ell=0$), which is obtained
from (\ref{Z4ell1Pfafffinal}) by omitting the Pfaffians. These
relations can also be used to now extract the massive spectral sum
rules in these instances.

\subsubsection*{\mbf{N_f=4\ell+3}}

The partition function (\ref{Z4equaloddgen}) for $n=2\ell+1$ is
evaluated identically as in (\ref{Z42ell1detexp}) with the obvious
change of integration domain and the substitutions
$\cos\theta^{~}_i\to2\kappa_i^2$ as above. The single unpaired
integration in this case can be expressed in terms of derivatives of
the function
\beq
\Xi_0(x)=\mbox{$\frac1{2\,\ii}$}~\sqrt{\mbox{$\frac\pi x$}}~
\erf\left(\,\ii\,\sqrt x\,\right) \ ,
\label{Xi0erf}\eeq
and we can thereby write the equal mass finite volume partition
function as the $(2\ell+2)\times(2\ell+2)$ Pfaffian formula
\bea
&\begin{tabular}{|l|}\hline\\
$\scriptstyle{Z_4^{(4\ell+3)}(\mu,\dots,\mu)}$\\
$\qquad\scriptstyle{=\e^{-(4\ell+1)\,\mu}~\pfaff\left[
\matrix{\scriptstyle\frac\ii{4\pi}~\int\limits_{-\infty}^\infty\dd q~
\frac{\Xi_q^{(i-1)}(4\mu)\,\Xi_q^{(j-1)}(4\mu)^*-
\Xi_q^{(j-1)}(4\mu)\,\Xi_q^{(i-1)}(4\mu)^*}{q+\ii\,\varepsilon}&\scriptstyle
\Xi_0^{(i-1)}(4\mu)\cr\scriptstyle-\Xi_0^{(j-1)}(4\mu)&\scriptstyle0}
\right]_{i,j=1,\dots,2\ell+1}}$\\$\scriptstyle{
\qquad\quad+~~(-1)^{N_\Dirac}~\bigl\{\mu\to-\mu\bigr\}}
$\\\\\hline\end{tabular}\nonu&
\label{Z4ell3Pfafffinal}\eea
This expression encompasses the case of three flavours ($\ell=0$) by
omitting the $(2\ell+1)\times(2\ell+1)$ blocks in
(\ref{Z4ell3Pfafffinal}). The Goldstone manifold (\ref{Goldstone4}) in
this case is diffeomorphic to a disk, $\mathcal{G}_4(3)\cong
SO(3)/SO(2)\times\zed_2\cong S^2/\zed_2 \cong D^2$, and the partition
function is equal to the function (\ref{Xi0erf}) evaluated at $x=4\mu$. This
simple evaluation follows of course from the explicit forms of the
angular integrals (\ref{Omega4expls}) in this case. The corresponding
massive spectral sum rule (\ref{sumrulegen}) is given by
\beq
\begin{tabular}{|l|}\hline\\
$\displaystyle\left.\left\langle\,\sum_{\zeta>0}
\frac1{\zeta^2+\mu^2}\right\rangle_4
{}~\right|_{N_f=3}=\frac{4\,\sqrt\mu~\e^{4\mu}-\sqrt\pi~{\rm erf}
\left(2\,\ii\,\sqrt\mu\,\right)}{12\,\sqrt\pi~\mu^2~{\rm erf}
\left(2\,\ii\,\sqrt\mu\,\right)}
$\\\\\hline\end{tabular}
\label{sumruleNf3beta1}\eeq

In addition to the sum rules, the expressions (\ref{kptexplbeta}),
(\ref{2flavourpart}), (\ref{Z4Pfafffinal}), (\ref{Z41Pfafffinal}),
(\ref{Z4ell1Pfafffinal}) and (\ref{Z4ell3Pfafffinal}) suffice to
determine the spectral $k$-point correlation functions in closed form
for an arbitrary number of flavours $N_f$ in the limit where the
collection of equal quark mass and (imaginary) Dirac operator
eigenvalues is completely degenerate. In particular, they may be used
to determine the correlators at the spectral origin for massless
quarks, as well as the smallest eigenvalue distribution functions
(\ref{holeprob},\ref{smallprob}).

\subsection*{Acknowledgments}

The author is grateful to G.~Akemann, P.~Damgaard, D.~Johnston and
G.~Vernizzi for helpful discussions and correspondence. This work was
supported in part by a PPARC Advanced Fellowship, by PPARC Grant
PPA/G/S/2002/00478, and by the EU-RTN Network Grant
MRTN-CT-2004-005104.


\begin{thebibliography}{99}

\baselineskip=12pt

\bibitem{GL1} J.~Gasser and H.~Leutwyler, Phys. Lett. {\bf B188}
  (1987) 477;\\
P.H.~Damgaard, Nucl. Phys. Proc. Suppl. {\bf 128} (2004) 47 [{\tt
  hep-lat/0310037}].

\bibitem{LS1} H.~Leutwyler and A.~Smilga, Phys. Rev. {\bf D46} (1992)
  5607.

\bibitem{BG1} C.~Bernard and M.F.L.~Golterman, Phys. Rev. {\bf D46}
  (1992) 853 [{\tt hep-lat/9204007}]; {\bf D49} (1994) 486 [{\tt
    hep-lat/9306005}];\\
M.F.L.~Golterman, Acta Phys. Pol. {\bf B25} (1994) 1731 [{\tt
  hep-lat/9411005}].

\bibitem{GHLW1} L.~Giusti, C.~Hoelbing, M.~L\"uscher and H.~Wittig,
  Comp. Phys. Commun. {\bf 153} (2003) 31 [{\tt hep-lat/0212012}];\\
L.~Giusti, M.~L\"uscher, P.~Weisz and H.~Wittig, J. High Energy
Phys. {\bf 0311} (2003) 023 [{\tt hep-lat/0309189}].

\bibitem{SV1} E.V.~Shuryak and J.J.M.~Verbaarschot, Nucl. Phys. {\bf
    A560} (1993) 306 [{\tt hep-th/9212088}];\\
J.J.M.~Verbaarschot and I.~Zahed, Phys. Rev. Lett. {\bf 70} (1993)
3852 [{\tt hep-th/9303012}].

\bibitem{ADMN1} G.~Akemann, P.H.~Damgaard, U.~Magnea and
  S.M.~Nishigaki, Nucl. Phys. {\bf B487} (1997) 721 [{\tt
    hep-th/9609174}].

\bibitem{Pisarski1} R.D.~Pisarski, Phys. Rev. {\bf D29} (1984) 2423.

\bibitem{VZ1} J.J.M.~Verbaarschot and I.~Zahed, Phys. Rev. Lett. {\bf
    73} (1994) 2288 [{\tt hep-th/9405005}].

\bibitem{HV1} M.A.~Halasz and J.J.M.~Verbaarschot, Phys. Rev. {\bf
    D52} (1995) 2563 [{\tt hep-th/9502096}].

\bibitem{Magnea1} U.~Magnea, Phys. Rev. {\bf D61} (2000) 056005 [{\tt
    hep-th/9907096}].

\bibitem{Magnea2} U.~Magnea, Phys. Rev. {\bf D62} (2000) 016005 [{\tt
    hep-th/9912207}].

\bibitem{NS1} T.~Nagao and K.~Slevin, J. Math. Phys. {\bf 34} (1993)
  2075.

\bibitem{DHKM1} P.H.~Damgaard, U.M.~Heller, A.~Krasnitz and T.~Madsen,
  Phys. Lett. {\bf B440} (1998) 129 [{\tt hep-th/9803012}].

\bibitem{DamNish} P.H. Damgaard and S.M. Nishigaki, Phys. Rev. {\bf
    D57} (1998) 5299 [{\tt hep-th/9711096}].

\bibitem{AD2} G.~Akemann and P.H.~Damgaard, Nucl. Phys. {\bf B598}
  (1998) 411 [{\tt hep-th/9801133}]; Phys. Lett. {\bf B432} (1998) 390
  [{\tt hep-th/9802174}].

\bibitem{Christ1} J.~Christiansen, Nucl. Phys. {\bf B547} (1999) 329
  [{\tt hep-th/9809194}].

\bibitem{HN1} C.~Hilmoine and R.~Niclasen, Phys. Rev. {\bf D62} (2000)
  096013 [{\tt hep-th/0004081}].

\bibitem{NN1} T.~Nagao and S.M.~Nishigaki, Phys. Rev. {\bf D63} (2001)
  045011 [{\tt hep-th/0005077}].

\bibitem{me} R.J. Szabo, Nucl. Phys. {\bf B598} (2001) 309 [{\tt
hep-th/0009237}].

\bibitem{ADDV1} G.~Akemann, D.~Dalmazi, P.H.~Damgaard and
  J.J.M.~Verbaarschot, Nucl. Phys. {\bf B601} (2001) 77 [{\tt
    hep-th/0011072}].

\bibitem{ADS1} T.~Andersson, P.H.~Damgaard and K.~Splittorff,
  Nucl. Phys. {\bf B707} (2005) 509 [{\tt hep-th/0410163}].

\bibitem{AD1} G. Akemann and P.H. Damgaard, Nucl. Phys. {\bf B576} (2000) 597
[{\tt hep-th/9910190}].

\bibitem{BMM1} H.W.~Braden, A.~Mironov and A.~Morozov,
  Phys. Lett. {\bf B514} (2001) 293 [{\tt hep-th/0105169}].

\bibitem{Zirn1} M.R.~Zirnbauer, J. Math. Phys. {\bf 37} (1996) 4986
  [{\tt math-ph/9808012}].

\bibitem{Helg1} S.~Helgason, {\it Differential Geometry, Lie Groups
    and Symmetric Spaces} (Academic Press, New York, 1978).

\bibitem{Helg2} S.~Helgason, {\it Groups and Geometric Analysis:
    Integral Geometry, Invariant Differential Operators and Spherical
    Functions} (Academic Press, New York, 1984).

\bibitem{CM1} M.~Caselle and U.~Magnea, Phys. Rept. {\bf 394} (2004)
  41 [{\tt cond-mat/0304363}].

\bibitem{Dam1} P.H.~Damgaard, Phys. Lett. {\bf B425} (1998) 151 [{\tt
    hep-th/9711047}].

\bibitem{Damsmall} S.M. Nishigaki, P.H.~Damgaard and T.~Wettig,
  Phys. Rev. {\bf D58} (1998) 087704 [{\tt hep-th/9803007}];\\
P.H.~Damgaard and S.M.~Nishigaki, Phys. Rev. {\bf
    D63} (2001) 045012 [{\tt hep-th/0006111}].

\bibitem{GW1} A.~Smilga and J.J.M.~Verbaarschot, Phys. Rev. {\bf D51}
  (1995) 829 [{\tt hep-th/9404031}];\\
A.D.~Jackson, M.K.~\c{S}ener and J.J.M.~Verbaarschot, Phys. Lett. {\bf
  B387} (1996) 355 [{\tt hep-th/9605183}];\\
T.~Guhr and T.~Wettig, Nucl. Phys. {\bf B506} (1997) 589 [{\tt
  hep-th/9704055}];\\
T.~Nagao and S.M.~Nishigaki, Phys. Rev. {\bf D62} (2000) 065006 [{\tt
  hep-th/0001137}].

\bibitem{Redlich1} A.N.~Redlich, Phys. Rev. Lett. {\bf 52} (1984) 18.

\bibitem{Mehta1} M.L.~Mehta, {\it Random Matrices} (Academic Press,
  San Diego, 1991).

\bibitem{Magnea3} U.~Magnea, Phys. Rev. {\bf D64} (2001) 018902 [{\tt
    hep-th/0009208}].

\bibitem{ZH1} M.R.~Zirnbauer and F.D.M.~Haldane, Phys. Rev. {\bf B52}
  (1995) 8729 [{\tt cond-mat/9504108}].

\bibitem{HC} Harish-Chandra, Am. J. Math. {\bf 79} (1957) 87.

\bibitem{mebook} R.J. Szabo, {\it Equivariant Cohomology and
    Localization of Path Integrals} (Springer-Verlag,
  Berlin-Heidelberg, 2000).

\bibitem{IZ} C. Itzykson and J.-B. Zuber, J. Math. Phys. {\bf 21}
  (1980) 411.

\bibitem{Bal1} A.B.~Balentekin and P.~Cassak, J. Math. Phys. {\bf 43}
  (2002) 604 [{\tt hep-th/0108130}].

\bibitem{BW1} B.~Schlittgen and T.~Wettig, J. Phys. {\bf A36} (2003)
  3195 [{\tt math-ph/0209030}].

\bibitem{NF1} T.~Nagao and P.J.~Forrester, Nucl. Phys. {\bf B509}
  (1998) 561.

\end{thebibliography}
\end{document}